\documentclass[11pt]{article}

\usepackage[margin=1in]{geometry}
\usepackage{amsmath,amssymb,bm}
\usepackage{graphicx}
\usepackage{booktabs}
\usepackage{array}
\usepackage{tabularx}
\usepackage{longtable}
\usepackage{multirow}
\usepackage{enumitem}
\usepackage{xcolor}
\usepackage{microtype}
\usepackage{tikz}
\usetikzlibrary{arrows.meta,positioning,fit,shapes.geometric,calc}
\usepackage[numbers,sort&compress]{natbib}
\usepackage{xurl}
\usepackage[colorlinks=true,linkcolor=blue,citecolor=blue,urlcolor=blue]{hyperref}
\usepackage{siunitx}
\usepackage{placeins}
\usepackage{authblk}
\usetikzlibrary{arrows.meta,positioning,calc}
\newcolumntype{Y}{>{\raggedright\arraybackslash}X}
\newcolumntype{P}[1]{>{\raggedright\arraybackslash}p{#1}}
\newcommand{\platform}{AIMBio-Mat}

\title{\platform{}: An AI-Native FAIR Platform for Closed-Loop Materials Discovery and Biomedical Translation}

\author[1,*]{D.-M. Mei}
\author[18]{K. Acharya}
\author[2]{C. M. Adhikari}
\author[1]{M. Adhikari}
\author[11]{S. Aryal}
\author[3]{B. V. Benson}
\author[11]{K. Bhatta}
\author[1]{S. Bhattarai}
\author[1]{N. Budhathoki}
\author[28]{A. M. Castillo}
\author[4]{D. Chakraborty}
\author[1]{S. Chhetri}
\author[5]{S. Choudhury}
\author[6]{T. A. Chowdhury}
\author[8]{R. D. Cruz}
\author[7]{B. Cui}
\author[11]{S. Dhital}
\author[1]{K.-M. Dong}
\author[9]{R. Gapuz}
\author[10]{A. Ghasemi}
\author[11]{E. Z. Gnimpieba}
\author[11]{B. D. S. Gurung}
\author[12]{H. A. Hashim}
\author[13]{R. I. Harry}
\author[14]{K.-E. Hasin}
\author[29]{M. K. Hassanzadeh}
\author[1]{M. K. Jha}
\author[1]{D. Kim}
\author[6]{K.-C. Kong}
\author[4]{B. Lama}
\author[15]{A. Mahat}
\author[11]{N. Maharjan}
\author[26]{A. Majeed}
\author[1]{J. Mammo}
\author[16]{M. M. Masud}
\author[17]{K. S. Moore}
\author[10]{A. Nawaz}
\author[18]{H. Oli}
\author[1]{S. A. Panamaldeniya}
\author[1]{L. Pandey}
\author[27]{R. Pandey}
\author[19]{Z. Peng}
\author[1]{A. Prem}
\author[11]{M. M. Rana}
\author[20]{K. Rana Magar}
\author[21]{R. Rizk}
\author[22]{C. S. Tadi}
\author[11]{L.-W. Wang}
\author[1]{Y. Yang}
\author[23]{G.-L. Yin}
\author[23]{C.-X. Yu}
\author[24]{D. Zeng}
\author[25]{M. Zhou}
\author[6]{Q. Zhou}

\affil[1]{Department of Physics, University of South Dakota, Vermillion, SD 57069, USA}
\affil[2]{Department of Chemistry, Physics and Materials Science, Fayetteville State University, Fayetteville, NC 28301, USA}
\affil[3]{PROMISE Lab, Sanford Research, Sioux Falls, SD 57104, USA}
\affil[4]{Department of Physics and Astronomy, University of Nebraska at Kearney, Kearney, NE 68849, USA}
\affil[5]{Department of Mechanical Engineering, University of Mississippi, University, MS 38677, USA}
\affil[6]{Department of Physics and Astronomy, University of Kansas, Lawrence, KS 66045, USA}
\affil[7]{Department of Mechanical and Materials Engineering, University of Nebraska--Lincoln, Lincoln, NE 68588, USA}
\affil[8]{Tiospa Zina Tribal School, Agency Village, SD 57262, USA}
\affil[9]{Dakota State University, Madison, SD 57042, USA}
\affil[10]{Department of Physics, University of Cincinnati, Cincinnati, OH 45221, USA}
\affil[11]{University of South Dakota, Vermillion, SD 57069, USA}
\affil[12]{Department of Physics, University of Central Florida, Orlando, FL 32816, USA}
\affil[13]{Department of Engineering and Applied Sciences, RAND Corporation, Santa Monica, CA 90401, USA}
\affil[14]{University of California, Merced, CA 95343, USA}
\affil[15]{School of Computing, Wichita State University, Wichita, KS 67260, USA}
\affil[16]{University of Illinois Chicago, Chicago, IL 60607, USA}
\affil[17]{STEP Lab, Massachusetts Institute of Technology, Cambridge, MA 02139, USA}
\affil[18]{Department of Physics, South Dakota School of Mines and Technology, Rapid City, SD 57701, USA}
\affil[19]{Department of Computer Science, University of Montana, Missoula, MT 59812, USA}
\affil[20]{Georgia State University, Atlanta, GA 30303, USA}
\affil[21]{Department of Computer Science, University of South Dakota, Vermillion, SD 57069, USA}
\affil[22]{College of Science and Engineering, Saint Cloud State University, St. Cloud, MN 56301, USA}
\affil[23]{Department of Agricultural and Biosystems Engineering, Iowa State University, Ames, IA 50011, USA}
\affil[24]{Information Systems, Dakota State University, Madison, SD 57042, USA}
\affil[25]{Department of Chemical and Materials Engineering, New Mexico State University, Las Cruces, NM 88003, USA}
\affil[26]{Department of Mechanical Engineering, Villanova University, Villanova, PA 19085, USA}
\affil[27]{Department of Computer Science, University of Kentucky, Lexington, KY 40506, USA}
\affil[28]{Lower Brule Schools, Lower Brule, SD 57548, USA}
\affil[29]{Department of Physics, Virginia Tech University, Blacksburg, VA 24061, USA}
\affil[*]{Corresponding author: \href{mailto:Dongming.Mei@usd.edu}{Dongming.Mei@usd.edu}}

\date{\today}

\begin{document}

\maketitle
\vspace{-1.5em}

\begin{abstract}
Materials discovery and biomedical translation increasingly require models that can reason across composition, processing, structure, biological response, manufacturability, safety, and governance constraints. Existing materials and biomedical data ecosystems are powerful but remain poorly coupled for AI-guided discovery. Here we present \platform{}, a conceptual framework for an AI-native, FAIR, and governance-aware decision layer that links materials provenance, biomedical context, knowledge graphs, uncertainty-aware machine learning, and human-in-the-loop active learning. The framework formulates biomedical-materials discovery as constrained multi-objective optimization under uncertainty and introduces practical requirements for metadata, model documentation, risk-tiered governance, evaluation metrics, and phased implementation. To make the roadmap testable, we add a minimum viable prototype specification and a worked pilot for AI-guided nanomaterials for drug delivery. \platform{} is positioned as exploratory and preclinical discovery infrastructure, not as clinical decision-support software; any clinical or regulated-device use would require separate validation, change control, and regulatory review. The central contribution is a publishable platform blueprint for converting fragmented materials and biomedical records into auditable, experimentally actionable, and translationally responsible discovery workflows.
\end{abstract}

\noindent\textbf{Keywords:} AI for science; materials informatics; biomedical innovation; FAIR data; active learning; knowledge graphs; biomaterials; drug delivery; explainable AI; digital twins; privacy-preserving learning.

\section{Introduction}

The discovery of functional materials and the development of biomedical technologies are foundational to progress in health care, energy, sustainability, advanced manufacturing, and national security. Traditional discovery pathways have relied on sequential experimentation, mechanistic modeling, expert intuition, and discipline-specific databases. These approaches remain indispensable, but they are increasingly insufficient for design spaces that combine chemical composition, synthesis route, processing history, microstructure, device architecture, biological response, toxicity, manufacturability, and regulatory constraints. A biomedical coating, for example, is not successful merely because it has a desirable modulus or surface energy; it must also be stable, biocompatible, manufacturable, sterilizable, and clinically safe. Similarly, a nanomaterial for drug delivery must satisfy multiple coupled constraints, including size, charge, degradation kinetics, immune response, drug loading, release profile, biodistribution, and toxicity.

In this manuscript, \emph{AI-native} means that data structures, metadata, model registries, uncertainty estimates, governance checks, and experiment-selection logic are designed into the platform from the beginning rather than appended after data collection. The intended near-term role of \platform{} is exploratory and preclinical scientific decision support: it prioritizes candidate materials, identifies missing evidence, records human decisions, and guides validation experiments. It is not proposed here as autonomous clinical decision-support software or as a deployed medical device. If outputs are later used to support clinical decisions or regulated device development, the platform would require task-specific validation, locked or controlled model-change procedures, monitoring, and regulatory review.

Materials informatics has shown that data-driven methods can accelerate materials discovery when high-quality data, descriptors, and predictive models are available. The Materials Project demonstrated how high-throughput first-principles calculations can be organized as an open materials genome infrastructure for screening inorganic materials~\citep{Jain2013MaterialsProject}. The Open Quantum Materials Database and AFLOW initiatives further illustrate how computed thermodynamic and electronic-structure data can be made searchable at scale~\citep{Kirklin2015OQMD,Curtarolo2012AFLOW}. Machine-learning methods have subsequently been applied to property prediction, inverse design, literature mining, and accelerated screening across molecular and materials systems~\citep{Butler2018MLMaterials,SanchezLengeling2018InverseDesign,Tshitoyan2019Embeddings,Merchant2023GNoME}. Autonomous experimentation and self-driving laboratories now extend these ideas from virtual screening to closed-loop experimental optimization~\citep{Kusne2020CAMEO,Stach2021Autonomous,Tom2024SDL,Szymanski2023ALab}.

Biomedical research has undergone a parallel data and AI transformation. PubChem provides chemical structures, bioassays, and related molecular data at large scale~\citep{Kim2023PubChem}; GenBank provides a long-standing infrastructure for nucleotide sequence data~\citep{Sayers2025GenBank}; and AI models such as AlphaFold and AlphaFold~3 have transformed biological structure prediction and biomolecular interaction modeling~\citep{Jumper2021AlphaFold,Abramson2024AlphaFold3}. Machine learning has also become increasingly important in drug discovery, target identification, diagnostic imaging, and clinical decision support~\citep{Vamathevan2019DrugDiscovery,Topol2019Medicine,Esteva2019Healthcare}. However, biomedical data infrastructures and materials data infrastructures are typically optimized for different communities, vocabularies, file formats, access controls, and validation practices.

The motivation for this manuscript was strengthened by the well-attended \emph{Workshop for AI-Powered Materials Discovery at Great Plains}, held at the University of South Dakota from June 22--25, 2025, with funding support from the National Science Foundation EPSCoR Workshop Program~\citep{AIMaterialsWorkshop2025}. The workshop brought together more than 200 participants and highlighted a clear regional and national need for AI-enabled materials discovery platforms that can connect materials synthesis, characterization, computation, biomedical applications, data governance, and workforce training. Discussions at the workshop emphasized that progress in AI-powered materials discovery will require more than individual algorithms or isolated datasets. It will require interoperable data infrastructure, shared metadata standards, closed-loop experimental workflows, trustworthy models, and mechanisms for translating materials discoveries into deployable technologies, especially in application areas where materials performance and biological response are tightly coupled.

The resulting gap is not merely technical. It is scientific and translational. Materials databases often lack biological-response metadata; biomedical datasets often lack detailed materials provenance; AI models frequently optimize a single property while real translational decisions require multi-objective tradeoffs; and privacy, regulatory, or intellectual-property constraints may prevent simple aggregation of biomedical data. Motivated by these challenges and by the workshop discussions, this paper expands the white-paper vision for a unified AI platform and database into a publishable scientific framework. The goal is to specify how such a platform can automate materials discovery while supporting biomedical applications through rigorous data models, closed-loop optimization, uncertainty quantification, governance mechanisms, and application-specific validation.

The specific contribution of this paper is to define \platform{} as a \emph{decision-oriented interoperability layer} rather than another isolated repository. Four elements distinguish the framework from existing materials databases, biomedical repositories, electronic laboratory notebooks/laboratory information-management systems (ELN/LIMS), and generic AI dashboards. First, it couples materials provenance, biological assay context, uncertainty, negative results, and governance metadata in a single AI-ready record model. Second, it exposes cross-domain relationships through a knowledge layer that preserves native community standards while enabling materials--biomedical queries. Third, it converts prediction into auditable decision support through constrained multi-objective optimization, active learning, applicability-domain checks, and human review. Fourth, it defines what a minimum viable prototype and a first pilot must demonstrate before broader translation. The manuscript is therefore intentionally presented as a framework and roadmap, not as a claim that \platform{} is already deployed, clinically validated, or regulatory-cleared. Accordingly, this manuscript is best classified as a framework or roadmap article rather than an experimental validation study.

\section{Scientific Rationale and Gap Analysis}

The scientific rationale for \platform{} rests on four observations. First, materials and biomedical records are often individually useful but rarely interoperable at the level needed for machine-actionable learning. Second, translational decisions require coupled objectives rather than isolated property prediction. Third, high-value evidence is distributed across different fidelity levels, from computation and in vitro assays to device testing and clinical observation. Fourth, trustworthy AI requires governance, provenance, and documentation as part of the technical system, not as an administrative afterthought. These observations are consistent with FAIR data stewardship, the TRUST principles for digital repositories, the CARE Principles for Indigenous Data Governance, and community efforts to document provenance and reuse conditions~\citep{Wilkinson2016FAIR,Lin2020TRUST,Carroll2020CARE,Lebo2013PROVO}. Naming CARE explicitly is important because biomedical-materials data may intersect with Indigenous, Tribal, and community-governed knowledge contexts, where collective benefit, authority to control, responsibility, and ethics must complement conventional open-data and repository principles. The subsections below translate these observations into platform-level design gaps.

\subsection{Fragmented and Heterogeneous Data Ecosystems}
Materials and biomedical data are heterogeneous by construction. A single material record may include composition, synthesis conditions, processing history, structural descriptors, microscopy, spectroscopy, electrical properties, thermal properties, mechanical behavior, corrosion response, environmental stability, and device-level performance. A biomedical record may include molecular structure, bioassay response, toxicity, histology, imaging, genomic sequence, patient phenotype, clinical outcome, treatment history, and regulatory metadata. These data differ in units, uncertainty estimates, provenance, privacy status, and allowed reuse.

The FAIR principles--findability, accessibility, interoperability, and reusability--provide a widely accepted foundation for scientific data stewardship \citep{Wilkinson2016FAIR}. Yet FAIR compliance alone is not enough for AI-native discovery. AI-ready scientific data must also contain sufficient context for model training, validation, and audit. This includes negative results, failed syntheses, calibration history, sample lineage, measurement uncertainty, data-use limitations, and protocol versions. Dataset documentation practices such as datasheets for datasets and model-reporting practices such as model cards are therefore relevant not only to social or commercial AI, but also to scientific AI platforms where reproducibility, bias, and intended use must be explicit \citep{Gebru2021Datasheets,Mitchell2019ModelCards}.

A useful way to define the gap is to ask what information is lost when a materials record is exported into a biomedical workflow, or when a biomedical assay result is imported into a materials model. Biomedical data standards such as MIAME for microarray experiments and ISA-Tab for bioscience investigations show that the context of measurement is often as important as the measured value itself~\citep{Brazma2001MIAME,Sansone2012ISATab}. \platform{} adopts the same logic for biomedical materials: a value such as ``cell viability'' or ``particle size'' should not be interpreted without the assay protocol, exposure dose, medium, characterization method, uncertainty estimate, and sample lineage.

\begin{table}[htp!]
\centering
\caption{Examples of information lost when materials and biomedical data are exchanged without cross-domain metadata.}
\label{tab:data_loss}
\footnotesize
\begin{tabularx}{\textwidth}{p{2.6cm}YY}
\toprule
\textbf{Record type} & \textbf{Commonly missing context} & \textbf{Consequence for AI-guided discovery} \\
\midrule
Nanoparticle characterization & Medium, aggregation state, measurement method, time after dispersion, temperature, concentration & Apparent size or charge becomes non-transferable across assays and laboratories \\
Surface or coating record & Sterilization route, roughness, surface chemistry, aging, cleaning protocol, substrate condition & Biological response may be incorrectly attributed to composition rather than surface state \\
Cell or tissue assay & Cell line, passage number, dose, exposure time, endpoint definition, controls, replicate structure & Toxicity or efficacy labels become ambiguous and hard to compare \\
Computational prediction & Functional, basis set, force field, convergence criteria, training domain, uncertainty, model version & Computed descriptors may be treated as ground truth even when outside the applicability domain \\
Clinical or patient-linked metadata & Consent, de-identification method, phenotype definition, site, inclusion criteria, data-use limits & Data may be scientifically useful but legally or ethically unusable without governance metadata \\
\bottomrule
\end{tabularx}
\end{table}

\subsection{Siloed Models and Narrow Objective Functions}
Current AI models in materials science and biomedicine are often trained for narrow tasks, such as predicting formation energy, band gap, adsorption energy, toxicity, binding affinity, image class, or disease risk. These task-specific models are useful, but biomedical materials development is inherently multi-objective. A drug-delivery nanoparticle may need to optimize biocompatibility, degradability, controlled release, manufacturability, pharmacokinetics, and regulatory acceptability. A biomedical implant may need to balance stiffness, fatigue strength, corrosion resistance, wear behavior, osseointegration, immune response, and long-term durability. A photocatalytic material intended for environmental or biomedical interfaces must optimize optical absorption, catalytic efficiency, chemical stability, low toxicity, and scalability.

Single-task optimization can therefore lead to candidates that perform well on one property but fail in practice. The platform must support Pareto-aware and constraint-aware decision-making rather than simple property ranking. It must also recognize that model uncertainty is not a nuisance term but a decision variable: high uncertainty may indicate a risky prediction, a data gap, or a valuable experiment. Active learning and Bayesian optimization provide a natural mechanism for selecting experiments that balance exploitation of promising candidates and exploration of uncertain regions \citep{Lookman2019ActiveLearning,Talapatra2018BOMU,Kusne2020CAMEO}.

A common but fragile workaround is to collapse all objectives into a single weighted score,
\begin{equation}
    U_{\bm{\lambda}}(\bm{x})=
    \sum_{m=1}^{M}\lambda_m \tilde f_m(\bm{x})
    -\sum_{k=1}^{K}\eta_k \max\!\left[0,g_k(\bm{x})\right]
    -\rho\,\sigma^2(\bm{x}),
    \label{eq:scalarutility}
\end{equation}
where $\tilde f_m$ are normalized objectives, $g_k$ are constraint violations, $\sigma^2(\bm{x})$ is predictive uncertainty, and $\bm{\lambda}$, $\bm{\eta}$, and $\rho$ encode user preferences. Equation~\eqref{eq:scalarutility} is useful for dashboards and sensitivity studies, but it should not replace Pareto-front analysis. Different stakeholders may reasonably choose different weights; therefore the platform should preserve the underlying multi-objective structure and expose tradeoffs rather than hiding them behind a single rank.

\subsection{The Translational Gap Between Prediction and Deployment}
A unified platform should not end at prediction. It must support the full translational pathway from hypothesis generation to candidate selection, experimental validation, process optimization, regulatory documentation, and stakeholder adoption. This distinction is especially important in biomedical applications. Models that influence clinical or preclinical decisions require clear performance evidence, traceable training data, defined use conditions, and risk controls. The U.S. Food and Drug Administration (FDA) has emphasized a total-product-life-cycle perspective for AI/machine-learning software as a medical device and has issued guidance on predetermined change control plans for AI-enabled devices \citep{FDA2021AIMLSaMD,FDA2025PCCP}. Although a materials-discovery platform is not necessarily a medical device, the same principles--traceability, locked versus adaptive models, validation evidence, and change control--are useful for any platform that may influence biomedical translation.

A second translational requirement is evidence reporting. AI models used in health-related studies increasingly face expectations for transparent reporting of training data, eligibility criteria, validation cohorts, endpoints, and model use conditions. Reporting extensions such as CONSORT-AI and SPIRIT-AI emphasize that AI-enabled interventions must describe inputs, outputs, human interaction, error cases, and integration into a workflow~\citep{Liu2020CONSORTAI,Rivera2020SPIRITAI}. For \platform{}, analogous documentation is needed even at the preclinical stage: a model-generated materials recommendation should record what data were used, what constraints were applied, what uncertainty was estimated, which human decision-maker approved the experiment, and what evidence would be sufficient to advance or reject the candidate.

Table~\ref{tab:useboundary} clarifies the intended boundary of use. This boundary is important because a platform can be scientifically useful for materials prioritization without being appropriate for clinical recommendation. The evidence requirements and governance controls must increase as decisions move from exploratory screening toward patient-facing use.

\begin{table}[htp!]
\centering
\caption{Boundary of use for \platform{} and corresponding governance posture.}
\label{tab:useboundary}
\footnotesize
\begin{tabularx}{\textwidth}{p{2.9cm}YYY}
\toprule
\textbf{Use category} & \textbf{Example platform role} & \textbf{Decision status} & \textbf{Minimum governance and validation posture} \\
\midrule
Exploratory materials discovery & Rank candidate compositions, coatings, or nanoparticles for laboratory testing & Research prioritization only & Metadata completeness, uncertainty reporting, applicability-domain flag, human approval before experiments \\
Preclinical evidence generation & Select candidates for in vitro assays, stability tests, or animal-study planning & Research evidence support & Locked data version, model card, validation report, decision log, safety and ethics review as appropriate \\
Regulated device-development evidence & Generate or organize evidence that may support a medical-device material or process & Regulatory-adjacent support & Change control, traceable training data, reproducible workflow bundle, independent validation, regulatory consultation \\
Clinical decision support & Recommend a material, implant, formulation, or intervention for a patient or care pathway & Not within the scope of this manuscript & Separate clinical validation, clinical governance, monitoring, regulatory review, and compliance with applicable medical-device or clinical-decision-support requirements \\
\bottomrule
\end{tabularx}
\end{table}

\section{Relationship to Existing Infrastructures and Novelty}

A major design requirement for \platform{} is that it should complement, not replace, existing community resources. The novelty of the proposed framework is the cross-domain coupling of materials provenance, biological response, AI-ready metadata, uncertainty-aware closed-loop optimization, and biomedical governance. Table~\ref{tab:comparison} summarizes representative infrastructures and the gap that remains for biomedical materials translation.

\begin{table}[!htbp]
\centering
\caption{Comparison of existing resources with the proposed role of \platform{}. The aim is not to duplicate mature community databases, but to create an interoperability and decision layer that connects materials, molecular, biological, and translational evidence.}
\label{tab:comparison}
\footnotesize
\begin{tabular}{P{2.35cm}P{1.9cm}P{2.7cm}P{3.1cm}P{3.1cm}}
\toprule
\textbf{Resource or ecosystem} & \textbf{Primary domain} & \textbf{Major strength} & \textbf{Limitation for biomedical materials translation} & \textbf{Role of \platform{}} \\
\midrule
Materials Project, OQMD, AFLOW, NOMAD & Inorganic and computational materials & Computed structures, thermodynamics, electronic properties, searchable materials records & Limited biological endpoints, assay context, toxicity, manufacturability, and clinical constraints & Ingest structures and computed descriptors while linking them to biomedical assays, process history, and translational constraints \\
OPTIMADE & Materials-data interoperability & Common API and query language for materials databases & Materials-focused; does not by itself define biomedical semantics or governance & Use as a model for cross-database access while extending semantic mappings to biological and biomedical endpoints \\
PubChem and ChEMBL & Chemistry, bioassays, drug-like molecules & Molecular structures, bioactivity records, assay information, compound identifiers & Weak representation of bulk materials, synthesis route, microstructure, surface state, and specimen provenance & Connect molecular/bioactivity records to materials descriptors, surface chemistry, release kinetics, and toxicity endpoints \\
GenBank and biomedical ontologies such as UMLS & Genomics and biomedical terminology & Sequence records and controlled biomedical vocabularies & Not designed for materials optimization or process--structure--property relationships & Provide patient/context and terminology mappings for personalized or biology-aware materials design \\
eNanoMapper and nanosafety resources & Nanomaterial safety and ontology & Nanomaterial descriptors, safety data, ontology-driven integration & Focused mainly on nanomaterial risk data; limited closed-loop optimization and platform governance & Provide reusable nanosafety descriptors and toxicity endpoints for active-learning workflows \\
Electronic laboratory notebooks (ELNs), laboratory information-management systems (LIMS), and local laboratory systems & Laboratory provenance & Instrument logs, protocols, sample tracking, local workflow continuity & Often local, heterogeneous, spreadsheet-driven, and not AI-ready across institutions & Supply adapters, schema mapping, quality checks, and provenance-preserving ingestion \\
\bottomrule
\end{tabular}
\end{table}

Thus, the proposed contribution is not another isolated data repository. It is a decision-oriented integration layer that asks whether a candidate material is simultaneously promising, feasible, safe, interpretable, auditable, and worth the next experiment. This distinction is important because biomedical materials translation depends on the coupling of multiple evidence streams: composition and processing history, physical and chemical properties, biological context, model uncertainty, manufacturing feasibility, privacy status, and regulatory relevance.

The integration layer should also recognize the difference between \emph{data exchange} and \emph{scientific interoperability}. Exchange standards such as HL7 FHIR and DICOM are essential for clinical and imaging contexts, while research packaging standards such as RO-Crate and workflow standards such as the Common Workflow Language help preserve computational provenance and reproducibility~\citep{HL7FHIR,DicomStandard,SoilandReyes2022ROCrate,Amstutz2016CWL}. \platform{} should not force all data into one universal schema. Instead, it should use adapters and crosswalks that preserve native community standards while exposing the subset of information needed for cross-domain discovery.

To make this novelty operational, Table~\ref{tab:mvp} defines a minimum viable \platform{} prototype. The prototype is intentionally modest: it does not require a fully autonomous laboratory or broad clinical integration. It must, however, demonstrate that cross-domain records can be ingested, validated, queried, modeled, and used to justify a next experiment with a complete audit trail.

\begin{table}[htp!]
\centering
\caption{Minimum viable prototype for testing \platform{} as a biomedical-materials decision layer.}
\label{tab:mvp}
\footnotesize
\begin{tabularx}{\textwidth}{p{2.7cm}YYY}
\toprule
\textbf{Prototype component} & \textbf{Minimum function} & \textbf{Example implementation} & \textbf{Validation criterion} \\
\midrule
AI-ready record schema & Capture material identity, process history, assay context, uncertainty, negative results, and governance fields & JSON/CSV schema with controlled units and required-field checks & At least 90\% completeness for required pilot fields and explicit flags for missing values \\
Adapters and crosswalks & Preserve native data formats while exposing common discovery fields & Importers for spreadsheets, ELN/LIMS exports, DLS/zeta files, release assays, and cell-viability tables & Reproducible ingestion of a fixed pilot dataset with unit harmonization and identifier resolution \\
Knowledge layer & Connect samples, batches, processes, assays, endpoints, models, and decision records & Lightweight graph database or typed relational schema with graph export & Correct answers to predefined competency questions across material, assay, and governance fields \\
Model registry & Store baseline models, uncertainty calibration, applicability-domain reports, and model cards & Versioned scikit-learn/GP/PyTorch models with reproducible workflow metadata & External or hold-out validation report plus calibration metric before decision use \\
Human-in-the-loop active learning & Recommend a short list of next experiments under constraints & Constrained EHVI or cost-weighted acquisition function plus expert approval screen & Logged recommendation, expert decision, experiment result, and model update for each iteration \\
Governance and audit & Record data-use limits, human decisions, model versions, and change history & Role-based access, immutable decision log, datasheet, and model card & Complete audit trail for every platform-generated recommendation \\
\bottomrule
\end{tabularx}
\end{table}

\section{Design Principles for an AI-Native Materials-Biomedical Platform}
The proposed platform, denoted \platform{}, is organized around eight design principles. Table~\ref{tab:principles} restates the principles as implementation requirements so that the manuscript reads as a platform blueprint rather than as a general AI vision.

\begin{table}[htp!]
\centering
\caption{Design principles and implementation requirements for \platform{}.}
\label{tab:principles}
\footnotesize
\begin{tabular}{P{2.7cm}P{4.0cm}P{6.1cm}}
\toprule
\textbf{Principle} & \textbf{Scientific rationale} & \textbf{Implementation requirement} \\
\midrule
AI-ready FAIR data & FAIR data are necessary but insufficient for model training and audit \citep{Wilkinson2016FAIR}. & Persistent identifiers, standardized units, uncertainty fields, machine-readable licenses, negative-result capture, and complete sample lineage. \\
Cross-domain semantic interoperability & Materials, biological, clinical, and regulatory concepts use different vocabularies. & Ontology mappings among materials descriptors, assay endpoints, UMLS/biomedical terminology, nanosafety descriptors, and process metadata \citep{Bodenreider2004UMLS,Hastings2015eNanoMapperOntology}. \\
Closed-loop discovery & Static databases do not prioritize the next experiment. & Human-in-the-loop or automated design--build--test--learn workflows with logged decisions and model updates \citep{Stach2021Autonomous,Tom2024SDL}. \\
Uncertainty-aware prediction & Biomedical decisions require confidence, applicability domain, and calibration, not only point predictions. & Prediction intervals, calibration diagnostics, out-of-distribution detection, and acquisition functions that account for uncertainty. \\
Mechanistic consistency & Purely data-driven correlations may fail outside the training domain. & Physical, chemical, biological, and clinical constraints embedded in feature engineering, model selection, or validation. \\
Interpretability and auditability & Translational decisions must be reviewable by scientists, clinicians, and regulators. & Model cards, explanation reports, feature-stability checks, and preference for interpretable models when adequate \citep{Mitchell2019ModelCards,Rudin2019Interpretable}. \\
Privacy-preserving governance & Biomedical data may include protected health information and data-use restrictions. & Role-based access, encryption, de-identification, audit logs, data-use agreements, and federated analysis when centralization is inappropriate \citep{HHSPrivacyRule,EuropeanUnion2016GDPR,Rieke2020Federated}. \\
Sustainable deployment & A platform must survive changes in datasets, models, instruments, and regulations. & Modular APIs, versioned schemas, workflow containers, cloud/HPC interoperability, community governance, and cost monitoring. \\
\bottomrule
\end{tabular}
\end{table}

The principles in Table~\ref{tab:principles} can also be interpreted as readiness gates. A dataset is not ready for closed-loop use until it has identifiers, provenance, uncertainty, and use restrictions. A model is not ready for recommendation until it has a validation report, an applicability domain, and a model card. A workflow is not ready for translation until it has a decision log, human review points, and a governance pathway. This gate-based interpretation prevents the platform from treating data availability as equivalent to scientific readiness.

\section{Platform Architecture}
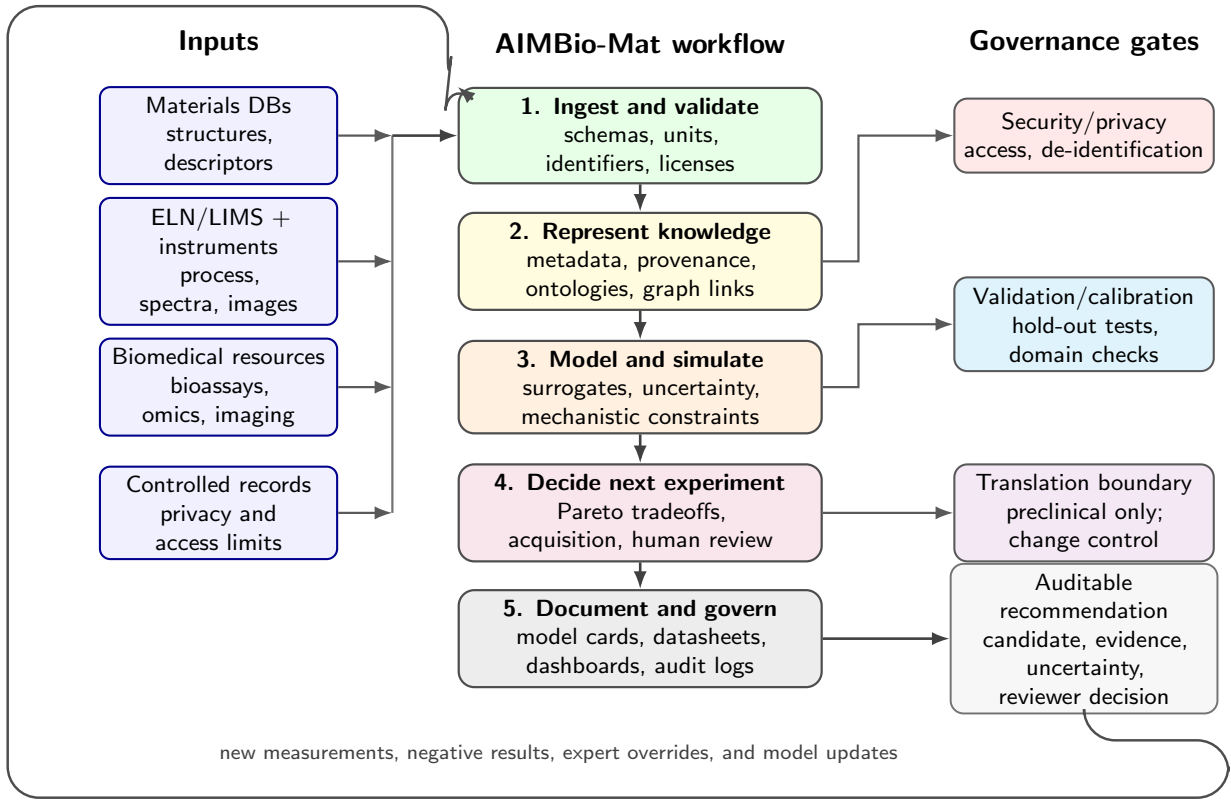
\begin{figure}[!htbp]
\centering
\resizebox{0.98\textwidth}{!}{%
\begin{tikzpicture}[
    x=1cm,y=1cm,
    font=\sffamily\footnotesize,
    >={Latex[length=2.4mm]},
    source/.style={
        rectangle, rounded corners=4pt,
        draw=blue!55!black, line width=0.8pt,
        fill=blue!6, align=center,
        text width=2.95cm, minimum height=0.95cm
    },
    layer/.style={
        rectangle, rounded corners=6pt,
        draw=black!70, line width=0.85pt,
        fill=#1, align=center,
        text width=4.65cm, minimum height=1.04cm
    },
    gate/.style={
        rectangle, rounded corners=5pt,
        draw=black!65, line width=0.8pt,
        fill=#1, align=center,
        text width=3.25cm, minimum height=1.00cm
    },
    note/.style={
        rectangle, rounded corners=5pt,
        draw=black!55, line width=0.75pt,
        fill=gray!6, align=center,
        text width=3.35cm, minimum height=1.04cm
    },
    mainarrow/.style={->, line width=1.0pt, draw=black!75},
    sidearrow/.style={->, line width=0.85pt, draw=black!65},
    busline/.style={line width=0.85pt, draw=black!55},
    feedback/.style={->, line width=0.95pt, draw=black!70, rounded corners=12pt}
]

\node[font=\sffamily\bfseries] at (0,1.25) {Inputs};
\node[font=\sffamily\bfseries] at (5.7,1.25) {AIMBio-Mat workflow};
\node[font=\sffamily\bfseries] at (11.7,1.25) {Governance gates};

\node[source] (matdb) at (0,0.00) {Materials DBs\\structures, descriptors};
\node[source] (lab)   at (0,-1.70) {ELN/LIMS + instruments\\process, spectra, images};
\node[source] (bio)   at (0,-3.40) {Biomedical resources\\bioassays, omics, imaging};
\node[source] (ctrl)  at (0,-5.10) {Controlled records\\privacy and access limits};

\node[layer=green!10]  (ingest)    at (5.7,0.00)   {\textbf{1. Ingest and validate}\\schemas, units, identifiers, licenses};
\node[layer=yellow!15] (knowledge) at (5.7,-1.70)  {\textbf{2. Represent knowledge}\\metadata, provenance, ontologies, graph links};
\node[layer=orange!12] (model)     at (5.7,-3.40)  {\textbf{3. Model and simulate}\\surrogates, uncertainty, mechanistic constraints};
\node[layer=purple!10] (decision)  at (5.7,-5.10)  {\textbf{4. Decide next experiment}\\Pareto tradeoffs, acquisition, human review};
\node[layer=gray!14]   (deploy)    at (5.7,-6.80)  {\textbf{5. Document and govern}\\model cards, datasheets, dashboards, audit logs};

\node[gate=red!9]    (privacy)     at (11.7,0.00)   {Security/privacy\\access, de-identification};
\node[gate=cyan!11]  (validation)  at (11.7,-2.55)  {Validation/calibration\\hold-out tests, domain checks};
\node[gate=violet!9] (translation) at (11.7,-5.10)  {Translation boundary\\preclinical only; change control};
\node[note]          (output)      at (11.7,-6.80)  {Auditable recommendation\\candidate, evidence, uncertainty, reviewer decision};

\coordinate (busTop)    at (2.35,0.00);
\coordinate (busLab)    at (2.35,-1.70);
\coordinate (busBio)    at (2.35,-3.40);
\coordinate (busBottom) at (2.35,-5.10);

\draw[sidearrow] (matdb.east) -- (busTop);
\draw[sidearrow] (lab.east)   -- (busLab);
\draw[sidearrow] (bio.east)   -- (busBio);
\draw[sidearrow] (ctrl.east)  -- (busBottom);

\draw[busline] (busBottom) -- (busTop);
\draw[mainarrow] (busTop) -- (ingest.west);

\draw[mainarrow] (ingest.south)    -- (knowledge.north);
\draw[mainarrow] (knowledge.south) -- (model.north);
\draw[mainarrow] (model.south)     -- (decision.north);
\draw[mainarrow] (decision.south)  -- (deploy.north);

\coordinate (kgate) at (8.65,-1.70);
\coordinate (mgate) at (8.65,-3.40);
\coordinate (dgate) at (8.65,-5.10);

\draw[sidearrow]
    (knowledge.east) -- (kgate) |- (privacy.west);

\draw[sidearrow]
    (model.east) -- (mgate) |- (validation.west);

\draw[sidearrow]
    (decision.east) -- (dgate) -- (translation.west);

\draw[mainarrow]
    (deploy.east) -- (output.west);

\node[
    align=center,
    font=\sffamily\scriptsize,
    text=black!75,
    fill=white,
    inner sep=2pt
] at (4.6,-8.35)
{new measurements, negative results, expert overrides, and model updates};

\coordinate (fbRight) at (13.65,-8.95);
\coordinate (fbLeft)  at (-2.85,-8.95);
\coordinate (fbUp)    at (-2.85,1.75);
\coordinate (fbEntryA) at (3.10,1.75);
\coordinate (fbEntryB) at (3.10,0.65);

\draw[feedback]
    (output.south) -- ++(0,-0.35)
    -| (fbRight)
    -- (fbLeft)
    -- (fbUp)
    -- (fbEntryA)
    -- (fbEntryB)
    -- ([xshift=-0.12cm,yshift=0.10cm]ingest.north west)
    -- (ingest.north west);

\end{tikzpicture}}
\caption{Publication-oriented architecture of \platform{}. Heterogeneous materials and biomedical inputs are validated, mapped into an AI-ready knowledge layer, modeled with uncertainty and mechanistic constraints, and converted into human-reviewed experimental recommendations. Governance gates operate across the workflow rather than as after-the-fact compliance checks.}
\label{fig:architecture1}
\end{figure}

Figure~\ref{fig:architecture1} shows the proposed architecture. The design separates the platform into five interacting layers: data ingestion, data and knowledge representation, AI and simulation, closed-loop discovery, and governance/user interaction. This separation is important because each layer has different technical requirements and risk controls.

The architecture is deliberately modular. Public data resources, local electronic laboratory notebooks (ELNs), laboratory information-management systems (LIMS), imaging repositories, and protected biomedical records should connect through adapters rather than through wholesale data migration. Each adapter should perform schema validation, unit harmonization, identifier mapping, license checking, and provenance capture before records enter the knowledge layer. In this sense, \platform{} functions as a governed integration fabric rather than a monolithic database.

\subsection{Data Ingestion Layer}
The ingestion layer connects external and internal data sources. Materials inputs may include computational databases, laboratory information-management systems, synthesis protocols, processing logs, microscopy images, diffraction data, spectroscopy, mechanical tests, and device measurements. Biomedical inputs may include chemical structures, bioassays, omics data, imaging, histology, patient-derived metadata, and preclinical or clinical outcomes. Public resources such as the Materials Project, OQMD, PubChem, and GenBank illustrate the value of structured data access, while interoperability efforts such as OPTIMADE show how standardized APIs can reduce friction across materials databases \citep{Jain2013MaterialsProject,Kirklin2015OQMD,Kim2023PubChem,Sayers2025GenBank,Andersen2021OPTIMADE}.

Ingestion should include automated validation checks: required fields, unit conversion, schema compliance, file integrity, license compatibility, quality flags, and privacy classification. Importantly, the platform should treat missing values and negative results as scientifically meaningful rather than as records to discard. In materials discovery, failed syntheses can define infeasible regions of composition--process space; in biomedical translation, adverse responses and null results are essential for safety-aware learning.

A practical ingestion pipeline should therefore include three queues: \emph{accepted records}, which satisfy minimum metadata requirements; \emph{curation-needed records}, which are scientifically valuable but incomplete; and \emph{restricted records}, which require privacy, intellectual-property, or regulatory review. This triage prevents high-value data from being lost while also preventing incomplete or restricted data from silently entering model training.

\subsection{Data and Knowledge Layer}
The data and knowledge layer stores curated records and connects them through a knowledge graph. A minimal record for a biomedical material should include material identity, batch or specimen identity, synthesis and processing descriptors, characterization methods, property values with uncertainty, biological endpoint definitions, assay conditions, provenance, access rights, and links to raw data. Knowledge graphs are useful because they can represent many-to-many relationships among materials, processes, biological targets, diseases, devices, and publications. They also support reasoning over partially connected data, such as linking a polymer chemistry to a surface functional group, a surface group to protein adsorption, and protein adsorption to immune response.

Formally, the knowledge layer can be represented as a typed graph
\begin{equation}
    \mathcal{G}=(\mathcal{V},\mathcal{E},\tau,\phi),
    \label{eq:knowledgegraph}
\end{equation}
where $\mathcal{V}$ contains entities such as materials, samples, processes, assays, biological targets, instruments, publications, and model versions; $\mathcal{E}$ contains typed relationships; $\tau$ maps nodes and edges to controlled vocabularies or ontologies; and $\phi$ stores attributes such as units, uncertainty, and access level. Equation~\eqref{eq:knowledgegraph} makes explicit that the platform must preserve relationships, not only tables of features.

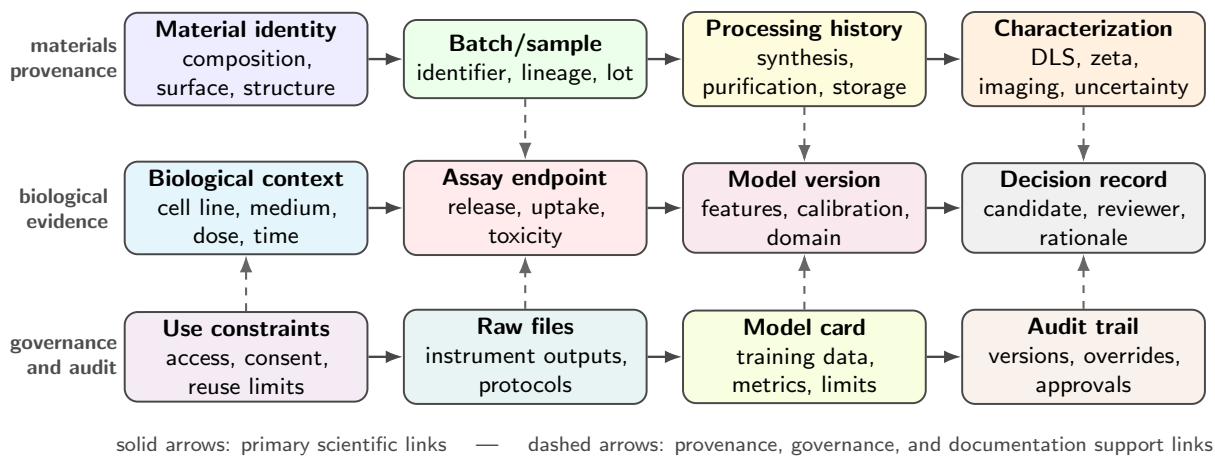
\begin{figure}[!htbp]
\centering
\resizebox{0.98\textwidth}{!}{%
\begin{tikzpicture}[
    x=1cm,y=1cm,
    font=\sffamily\footnotesize,
    >={Latex[length=2.3mm]},
    entity/.style={rectangle, rounded corners=5pt, draw=black!72, line width=0.75pt, fill=#1, align=center, text width=3.05cm, minimum height=1.05cm, inner sep=3pt},
    rowtag/.style={font=\sffamily\scriptsize\bfseries, text=black!70, align=right},
    edge/.style={->, line width=0.90pt, draw=black!72, rounded corners=6pt},
    support/.style={->, dashed, line width=0.80pt, draw=black!60, rounded corners=6pt}
]

\node[rowtag] at (-2.45, 1.65) {materials\\provenance};
\node[rowtag] at (-2.45,-0.35) {biological\\evidence};
\node[rowtag] at (-2.45,-2.35) {governance\\and audit};

\node[entity=blue!7]    (material) at (0.0, 1.65) {\textbf{Material identity}\\composition,\\surface, structure};
\node[entity=green!8]   (batch)    at (3.75, 1.65) {\textbf{Batch/sample}\\identifier, lineage, lot};
\node[entity=yellow!16] (process)  at (7.50, 1.65) {\textbf{Processing history}\\synthesis,\\purification, storage};
\node[entity=orange!12] (char)     at (11.25, 1.65) {\textbf{Characterization}\\DLS, zeta,\\imaging, uncertainty};

\node[entity=cyan!9]    (bioctx)   at (0.0,-0.35) {\textbf{Biological context}\\cell line, medium, dose, time};
\node[entity=red!8]     (assay)    at (3.75,-0.35) {\textbf{Assay endpoint}\\release, uptake,\\toxicity};
\node[entity=purple!9]  (model)    at (7.50,-0.35) {\textbf{Model version}\\features, calibration,\\domain};
\node[entity=gray!12]   (decision) at (11.25,-0.35) {\textbf{Decision record}\\candidate, reviewer,\\rationale};

\node[entity=violet!8]  (govern)   at (0.0,-2.35) {\textbf{Use constraints}\\access, consent,\\reuse limits};
\node[entity=teal!9]    (raw)      at (3.75,-2.35) {\textbf{Raw files}\\instrument outputs,\\protocols};
\node[entity=lime!12]   (modelcard)at (7.50,-2.35) {\textbf{Model card}\\training data,\\metrics, limits};
\node[entity=brown!10]  (auditlog) at (11.25,-2.35) {\textbf{Audit trail}\\versions, overrides,\\approvals};

\draw[edge] (material.east) -- (batch.west);
\draw[edge] (batch.east) -- (process.west);
\draw[edge] (process.east) -- (char.west);

\draw[edge] (bioctx.east) -- (assay.west);
\draw[edge] (assay.east) -- (model.west);
\draw[edge] (model.east) -- (decision.west);

\draw[edge] (govern.east) -- (raw.west);
\draw[edge] (raw.east) -- (modelcard.west);
\draw[edge] (modelcard.east) -- (auditlog.west);

\draw[support] (govern.north) -- (bioctx.south);
\draw[support] (batch.south) -- (assay.north);
\draw[support] (raw.north) -- (assay.south);
\draw[support] (process.south) -- (model.north);
\draw[support] (modelcard.north) -- (model.south);
\draw[support] (char.south) -- (decision.north);
\draw[support] (auditlog.north) -- (decision.south);

\node[align=center, font=\sffamily\scriptsize, text=black!72] at (5.63,-3.55)
{solid arrows: primary scientific links \quad | \quad dashed arrows: provenance, governance, and documentation support links};

\end{tikzpicture}}
\caption{Publication-oriented knowledge-graph structure for \platform{}. The diagram separates materials provenance, biological evidence, and governance/audit artifacts into clean lanes. Solid arrows show primary scientific relationships, while dashed arrows show supporting provenance, documentation, and use-constraint links. This layout avoids arrow crossings and avoids placing relationship text between boxes.}
\label{fig:knowledgegraph}
\end{figure}

Figure~\ref{fig:knowledgegraph} illustrates the minimum graph structure needed for biomedical-materials discovery. The point is not to force all records into one universal table, but to retain typed links among materials, processing history, biological context, model versions, governance limits, and decision records.

\subsection{AI and Simulation Layer}
The AI and simulation layer supports multiple model families. Graph neural networks can represent molecules, crystalline structures, and materials graphs; transformer and embedding models can mine literature and protocol text; surrogate models can emulate expensive simulations; and physics-informed models can enforce conservation laws, thermodynamic constraints, or mechanistic relationships. For biomedical applications, predictive models must be paired with careful validation and domain limits because the cost of erroneous predictions may be high.

The layer should also support model documentation. Each model should have a machine-readable model card describing training data, intended use, excluded use, performance, uncertainty calibration, bias assessment, version history, and known limitations \citep{Mitchell2019ModelCards}. For high-stakes applications, interpretability must be designed into the modeling strategy rather than appended as an afterthought \citep{Rudin2019Interpretable}. Post hoc explanation methods such as LIME and SHAP can still be useful for exploratory analysis, but they should not substitute for validation, calibration, and mechanistic scrutiny \citep{Ribeiro2016LIME,Lundberg2017SHAP}.

The model layer should therefore maintain a model registry rather than a single best model. For each prediction task, the registry should store baseline models, interpretable models, high-capacity models, physics-informed variants, and calibrated uncertainty models. Model comparison should include external validation, calibration, subgroup or context analysis, and out-of-distribution testing. This is especially important when a model trained on one assay, laboratory, or material class is used to recommend experiments in another.

\subsection{Closed-Loop Discovery Layer}
\begin{figure}[htp!]
\centering
\resizebox{0.82\textwidth}{!}{%
\begin{tikzpicture}[
    font=\small,
    block/.style={rectangle, rounded corners, draw=black!70, line width=0.7pt, minimum width=3.0cm, minimum height=1.0cm, align=center, fill=#1},
    arrow/.style={-{Latex[length=2.2mm]}, thick, black!75}
]
\node[block=blue!8] (design) {Design\\candidate generation};
\node[block=green!8, right=1.05cm of design] (build) {Build\\synthesis/fabrication};
\node[block=orange!10, right=1.05cm of build] (test) {Test\\characterization/assay};
\node[block=purple!9, below=1.0cm of build] (learn) {Learn\\model update};
\node[block=gray!12, below=1.0cm of test] (decide) {Decide\\rank/select/stop};

\draw[arrow] (design) -- (build);
\draw[arrow] (build) -- (test);
\draw[arrow] (test) -- (decide);
\draw[arrow] (decide) -- (learn);
\draw[arrow] (learn) -- (design);
\draw[arrow] (test.south) -- (learn.east);

\node[align=center, above=0.28cm of build] {\textbf{Closed-loop discovery cycle}};
\node[align=center, below=0.35cm of learn, text width=9.5cm] {Each iteration records provenance, uncertainty, failed trials, human decisions, and model/version changes.};
\end{tikzpicture}}
\caption{Design--build--test--learn cycle implemented by the closed-loop discovery layer. The system can operate as a human-in-the-loop workflow before full laboratory automation is available.}
\label{fig:closedloop}
\end{figure}
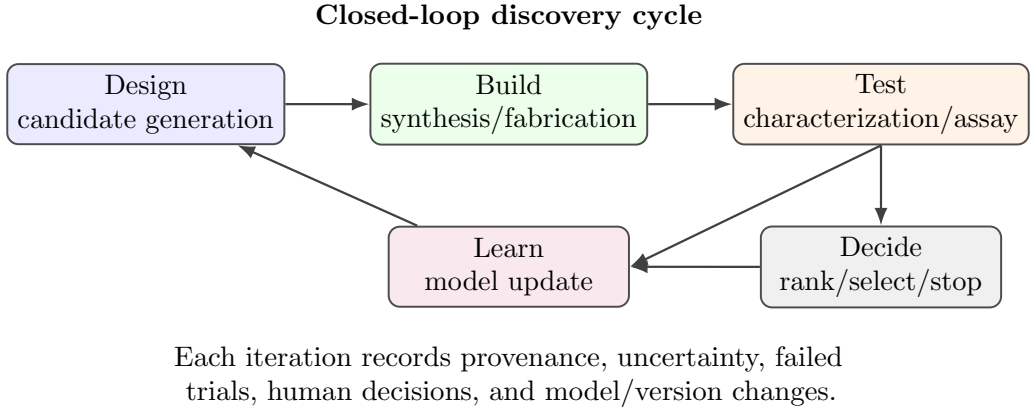

The closed-loop layer converts model outputs into decisions (Figure~\ref{fig:closedloop}). It ranks candidates, proposes experiments, updates models after new data are collected, and records the rationale for each decision. This layer is inspired by autonomous experimentation systems, but it does not require full robotic automation in its initial form. Many laboratories can benefit from ``human-in-the-loop'' active learning, where AI proposes a short list of high-value experiments and domain experts approve, revise, or reject the suggestions. At a decision point, the platform can recommend
\begin{equation}
    \bm{x}_{t+1} \in \arg\max_{\bm{x}\in \Omega_t}
    a_t(\bm{x}) \quad \text{subject to} \quad
    \bm{x}\in \mathcal{A}_{\mathrm{allowed}} \cap \mathcal{A}_{\mathrm{safe}},
    \label{eq:decisionrule}
\end{equation}
where $a_t$ is an acquisition function, $\Omega_t$ is the current feasible design space, $\mathcal{A}_{\mathrm{allowed}}$ encodes governance and access constraints, and $\mathcal{A}_{\mathrm{safe}}$ encodes safety or ethical exclusions. The human reviewer can accept the recommendation, override it, or request additional evidence; all three actions should be logged.

\subsection{Governance and User Layer}

The governance and user layer manages access, privacy, reproducibility, regulatory documentation, and collaboration. Biomedical data require special protection because they may contain electronic protected health information (ePHI), which refers to individually identifiable health information that is created, received, maintained, or transmitted electronically. In the United States, such information is regulated under the Health Insurance Portability and Accountability Act (HIPAA), particularly the HIPAA Privacy Rule, which governs the permissible use and disclosure of protected health information, and the HIPAA Security Rule, which establishes administrative, physical, and technical safeguards for electronic protected health information \citep{HHSPrivacyRule,HHSSecurityRule}. For data concerning persons in the European Union, the platform may also need to comply with the General Data Protection Regulation (GDPR), a comprehensive data-protection framework that governs the collection, processing, storage, transfer, and rights-based control of personal data \citep{EuropeanUnion2016GDPR}.

The platform should therefore implement role-based access control, encryption at rest and in transit, audit logs, data-use agreements, de-identification or tokenization when appropriate, consent and data-processing records, and review workflows for controlled data release. These governance mechanisms should be treated as core infrastructure rather than as after-the-fact compliance additions, especially when materials-discovery data are linked to biomedical assays, patient-derived samples, imaging data, clinical metadata, or translational validation studies.

The user layer should provide dashboards for data exploration, model training, experiment planning, validation review, and regulatory documentation. Researchers should be able to ask: Which candidates lie on the current Pareto front? Which features control model predictions? Which measurements would most reduce uncertainty? Which samples have incomplete provenance? Which model version generated a recommendation? Which datasets are governed by HIPAA, GDPR, data-use agreements, or controlled-access restrictions? These questions are essential for scientific trust, reproducibility, responsible collaboration, and translational readiness.
\section{Mathematical Formulation of Platform-Guided Discovery}

\subsection{Data Object and Provenance Representation}
A platform record can be represented as a structured object
\begin{equation}
    \mathcal{D}_i = \{\bm{x}_i, \bm{p}_i, \bm{s}_i, \bm{c}_i, \bm{y}_i, \bm{u}_i, \bm{g}_i, \bm{r}_i\},
    \label{eq:datarecord}
\end{equation}
where $\bm{x}_i$ denotes composition or molecular structure, $\bm{p}_i$ processing and synthesis variables, $\bm{s}_i$ structural and surface descriptors, $\bm{c}_i$ context such as assay or patient-relevant conditions, $\bm{y}_i$ measured outcomes, $\bm{u}_i$ uncertainty estimates, $\bm{g}_i$ governance metadata, and $\bm{r}_i$ provenance links to raw data, protocols, instruments, and model versions. Equation~\eqref{eq:datarecord} emphasizes that the learning object is not simply a feature vector; it is a traceable scientific record.

A simple AI-readiness score for a record can be defined as
\begin{equation}
    Q_i =
    \alpha M_i+\beta U_i+\gamma P_i+\delta G_i+\zeta N_i,
    \label{eq:readiness}
\end{equation}
where $M_i$ measures completeness of required metadata, $U_i$ measures uncertainty reporting, $P_i$ measures provenance depth, $G_i$ measures governance clarity, and $N_i$ indicates whether negative or null outcomes are recorded when relevant. The coefficients should be set by the pilot governance team. Records with low $Q_i$ may still be stored, but they should not be used for high-stakes model training without curation.

\subsection{Constrained Multi-Objective Optimization}
Materials-biosystem design can be expressed as a constrained multi-objective optimization problem:
\begin{equation}
    \bm{x}^{\ast} \in \arg\max_{\bm{x}\in\Omega} \; \bm{f}(\bm{x}) = \left[f_1(\bm{x}), f_2(\bm{x}), \ldots, f_M(\bm{x})\right]
    \quad \text{subject to} \quad g_k(\bm{x}) \leq 0,\; k=1,\ldots,K,
    \label{eq:multiobjective}
\end{equation}
where $\Omega$ is the feasible design space, $f_m$ are objectives such as efficacy, stability, biocompatibility, manufacturability, and cost, and $g_k$ are constraints such as toxicity, regulatory exclusion, processing limits, or data-use restrictions. In practice, Eq.~\eqref{eq:multiobjective} does not produce a single universal optimum; it produces a Pareto set from which candidates must be selected based on scientific and translational priorities.

\subsection{Uncertainty-Aware Acquisition Function}
At iteration $t$, the platform has data $\mathcal{D}_t$ and a probabilistic model that predicts a joint distribution over objectives and constraints. For multi-objective discovery, a natural acquisition criterion is expected hypervolume improvement (EHVI), which measures the expected increase in the dominated hypervolume of the Pareto set relative to a reference point. For noisy, parallel, and constrained settings, variants such as noisy expected hypervolume improvement provide a practical Bayesian-optimization foundation \citep{Daulton2021NEHVI}. In \platform{}, this idea can be written as a cost- and safety-weighted constrained acquisition function:
\begin{equation}
    a_t(\bm{x}) =
    \mathrm{EHVI}_t(\bm{x};\bm{r})
    \; P_t\!\left(\bigcap_{k=1}^{K}\{g_k(\bm{x}) \leq 0\}\right)
    \; \frac{S_t(\bm{x})}{C_t(\bm{x})+\epsilon},
    \label{eq:acquisition}
\end{equation}
where $\mathrm{EHVI}_t(\bm{x};\bm{r})$ is the expected hypervolume improvement relative to reference point $\bm{r}$, $P_t(\cap_k\{g_k(\bm{x})\leq0\})$ is the model-estimated probability that all feasibility, toxicity, manufacturing, and governance constraints are satisfied, $S_t(\bm{x})\in[0,1]$ is a safety or trust score, $C_t(\bm{x})$ is an estimated cost in time, money, sample availability, or instrument burden, and $\epsilon>0$ prevents singular behavior for low-cost experiments. Equation~\eqref{eq:acquisition} is intended as an operational template rather than a universal formula. Its purpose is to prevent the platform from selecting candidates solely because one predicted property is high; instead, the next experiment should offer expected Pareto improvement, acceptable feasibility, explicit safety, and manageable cost.

\subsection{Multi-Fidelity and Context-Aware Learning}
Materials--biomedical evidence rarely comes from one fidelity level. A single candidate may have density-functional-theory descriptors, molecular simulations, in vitro assays, animal-model data, manufacturing tests, and clinical observations. These data streams differ in cost, bias, uncertainty, and relevance. A simplified context-aware loss function can be expressed as
\begin{equation}
    \mathcal{L}(\theta) =
    \sum_{i=1}^{N} w_{\ell_i}
    \left\| \bm{y}_i-\hat{\bm{y}}_{\theta}
    (\bm{x}_i,\bm{p}_i,\bm{s}_i,\bm{c}_i,\ell_i) \right\|^2
    + \lambda \mathcal{R}(\theta),
    \label{eq:multifidelity}
\end{equation}
where $\ell_i$ denotes fidelity or evidence level, $w_{\ell_i}$ weights measurements according to reliability or decision relevance, and $\mathcal{R}(\theta)$ imposes regularization or mechanistic constraints. This formulation makes explicit that an assay label cannot be interpreted without context: dose, medium, exposure time, cell line, sterilization condition, and patient-relevant variables may change the meaning of the same measured endpoint.
\subsection{Model Updating and Audit Trail}
After a new experiment is performed, the data update from $\mathcal{D}_t$ to $\mathcal{D}_{t+1}=\mathcal{D}_t\cup\{\mathcal{D}_{\mathrm{new}}\}$. A Bayesian view writes model updating as
\begin{equation}
    p(\theta \mid \mathcal{D}_{t+1}) \propto p(\mathcal{D}_{\mathrm{new}} \mid \theta)\,p(\theta \mid \mathcal{D}_{t}),
    \label{eq:bayesupdate}
\end{equation}
where $\theta$ denotes model parameters. Even when the implemented model is not explicitly Bayesian, the platform should preserve an equivalent audit trail: data version, model version, training configuration, validation report, and decision log.

\section{Representative Data Resources and Interoperability Requirements}
Table~\ref{tab:resources} summarizes representative resources and standards that inform the proposed platform. The goal is not to replace these platforms, but to connect them through well-defined APIs, metadata mappings, domain-specific governance, and decision-oriented workflows.

\begin{table}[htp!]
\centering
\caption{Representative data resources, standards, and interoperability requirements relevant to an integrated materials--biomedical platform.}
\label{tab:resources}
\footnotesize
\begin{tabular}{P{2.6cm}P{2.1cm}P{3.0cm}P{5.4cm}}
\toprule
\textbf{Resource or standard} & \textbf{Primary domain} & \textbf{Integration value} & \textbf{Platform requirement} \\
\midrule
Materials Project \citep{Jain2013MaterialsProject} & Computed inorganic materials & Formation energies, structures, electronic properties, screening data & API ingestion, structure descriptors, calculation provenance, and uncertainty/functional metadata \\
OQMD, AFLOW, and NOMAD \citep{Kirklin2015OQMD,Curtarolo2012AFLOW,Draxl2019NOMAD} & High-throughput computational materials & Thermodynamic stability, electronic structure, and reusable computational records & Cross-database identifiers, metadata harmonization, workflow provenance, and duplicate-resolution logic \\
OPTIMADE \citep{Andersen2021OPTIMADE} & Materials-data interoperability & Common API for accessing materials databases & Adoption of common query syntax, schema mapping, and extensible endpoints for biomedical links \\
PubChem and ChEMBL \citep{Kim2023PubChem,Zdrazil2024ChEMBL} & Chemistry and bioassay data & Chemical structures, assays, identifiers, compound--target relationships, bioactivity data & Molecular descriptors, assay harmonization, toxicity endpoints, and links to materials-mediated delivery or surface chemistry \\
GenBank and AlphaFold resources \citep{Sayers2025GenBank,Jumper2021AlphaFold,Abramson2024AlphaFold3} & Genomics and biomolecular structure & Sequence records, BioProject/BioSample linkage, protein structures and biomolecular interactions & Controlled access when linked to sensitive metadata; mappings from biological targets to materials constraints \\
UMLS and clinical common-data models \citep{Bodenreider2004UMLS,Klann2019OMOP} & Biomedical terminology and clinical data & Concept mapping across clinical vocabularies and observational health-data models & Controlled terminology mapping, phenotype definitions, and separation of identifiable clinical records from materials workflows \\
HL7 FHIR and DICOM \citep{HL7FHIR,DicomStandard} & Clinical exchange and imaging & Standardized health-data exchange and medical-image representation & Controlled adapters for clinical metadata, imaging-derived descriptors, and separation of research use from care-delivery records \\
RO-Crate and Common Workflow Language \citep{SoilandReyes2022ROCrate,Amstutz2016CWL} & Research packaging and workflows & Machine-readable packaging of data, software, workflows, and provenance & Reproducible workflow bundles for model training, simulation, and experiment-planning pipelines \\
eNanoMapper \citep{Jeliazkova2015eNanoMapper,Hastings2015eNanoMapperOntology} & Nanomaterial safety and ontology & Nanomaterial descriptors, nanosafety endpoints, and ontology-driven integration & Reuse of nanomaterial descriptors, toxicity labels, exposure context, and safety metadata \\
Datasheets and model cards \citep{Gebru2021Datasheets,Mitchell2019ModelCards} & AI documentation & Transparency for datasets and models & Required documentation for data/model release, intended use, limitations, and internal audit \\
\bottomrule
\end{tabular}
\end{table}

The most important interoperability requirement is preservation of scientific meaning across domains. For example, a nanoparticle size measurement must specify the measurement method, medium, aggregation state, temperature, concentration, and uncertainty. A toxicity label must specify cell line or organism, exposure time, dose, assay endpoint, controls, and statistical confidence. Without this context, the platform risks producing models that are numerically accurate on a benchmark but scientifically unreliable in deployment.

Interoperability should therefore be validated through competency questions rather than only through schema compliance. Examples include: Which polymer coatings have comparable sterilization histories and cell-adhesion assays? Which nanoparticles have release profiles measured in physiologically relevant media and toxicity assays with matched dose ranges? Which candidate materials have both a computed stability descriptor and a biological-response endpoint with sufficient provenance? If the platform cannot answer such questions reproducibly, the data are not yet interoperable for discovery.

\section{Representative Scientific Use Cases}
The following use cases are intended as technically grounded pilot workflows rather than claims of completed validation. Each use case can be evaluated against an expert-only, random-search, or single-objective baseline. Table~\ref{tab:usecases} defines the data inputs, model outputs, validation steps, and success metrics that would make each pilot testable.

\begin{table}[htp!]
\centering
\caption{Concrete pilot specifications for representative \platform{} use cases.}
\label{tab:usecases}
\footnotesize
\begin{tabular}{P{2.35cm}P{3.1cm}P{2.65cm}P{2.55cm}P{2.8cm}}
\toprule
\textbf{Use case} & \textbf{Input data} & \textbf{Model outputs} & \textbf{Validation} & \textbf{Example success metric} \\
\midrule
Nanomaterials for drug delivery & Size, zeta potential, morphology, ligand density, polymer or inorganic chemistry, drug properties, release media, dose, cell line, toxicity, uptake, release kinetics & Release half-life, cytotoxicity probability, cellular uptake, manufacturability score, feasibility constraints & In vitro release assay, cell viability, uptake imaging, stability testing, limited pharmacokinetic follow-up & Fewer experiments to identify candidates satisfying release, toxicity, and manufacturing constraints \\
Biomedical implants & Alloy or polymer composition, processing route, porosity, surface treatment, topography, coating chemistry, mechanical tests, corrosion, biological response & Fatigue/wear risk, corrosion resistance, osseointegration likelihood, immune-response risk, sterilization compatibility & Mechanical fatigue/wear tests, corrosion assays, cell adhesion/proliferation, animal or clinical registry linkage when appropriate & Reduction in failed screening cycles relative to expert-only candidate selection \\
Photocatalytic{/} antimicrobial surfaces & Band gap, band-edge positions, defect descriptors, surface adsorption energies, synthesis variables, illumination, medium, pollutant or microbial endpoint, cytotoxicity & Photocatalytic rate, stability, toxicity risk, antimicrobial activity, environmental compatibility & Photodegradation assay, antimicrobial test, cytotoxicity screen, stability under repeated cycling & Higher activity-to-toxicity ratio at lower experimental cost \\
Personalized biomaterials & Materials descriptors, patient-relevant constraints, anatomy, phenotype/genotype-derived risk categories, immune status, prior treatment, controlled clinical metadata & Patient- or subgroup-specific material constraints, risk scores, candidate ranking under privacy controls & Federated or controlled-access validation, retrospective registry analysis, subgroup performance testing & Improved subgroup-specific performance without centralizing protected health information \\
\bottomrule
\end{tabular}
\end{table}

The four use cases also differ in acceptable risk. Environmental photocatalysis may tolerate early screening of many low-cost candidates, whereas patient-specific biomaterials require stricter governance and validation before any clinical interpretation. The platform should therefore attach a risk tier to each workflow and require stronger evidence as the consequence of a wrong recommendation increases.

\subsection{AI-Guided Nanomaterials for Drug Delivery}
Drug delivery is used here as the primary worked example because it naturally couples materials descriptors, molecular properties, release behavior, biological response, and translational constraints. Nanoparticles can be engineered to modulate circulation, targeting, degradation, and release, but their performance is context dependent: size distribution, surface chemistry, ligand density, protein-corona formation, medium, dose, and cell model can change the interpretation of the same endpoint. Public chemistry and bioassay resources such as PubChem and ChEMBL can provide molecular and activity context, while nanosafety resources such as eNanoMapper can provide descriptors and safety-relevant endpoints for engineered nanomaterials \citep{Kim2023PubChem,Zdrazil2024ChEMBL,Jeliazkova2015eNanoMapper}.

Rather than developing a second, overlapping narrative here, the operational details are folded into the worked pilot below. The pilot defines the design variables, required record structure, closed-loop decision process, and success criteria needed to make a drug-delivery nanoparticle workflow testable rather than merely illustrative.

\subsection{Worked Pilot: Constrained Active Learning for Drug-Delivery Nanoparticles}
\label{subsec:workedpilot}

To make the framework testable, an initial pilot can focus on polymeric or hybrid nanoparticles designed to deliver a small-molecule therapeutic in a defined in vitro model. The design vector can be formalized as
\begin{equation}
    \bm{x}=\left(x_{\mathrm{poly}},\; r_{\mathrm{core/drug}},\; \ell_{\mathrm{lig}},\; f_{\mathrm{PEG}},\; x_{\mathrm{solv}},\; \omega_{\mathrm{mix}},\; x_{\mathrm{pur}}
    \right),
    \label{eq:nanoparticle_design}
\end{equation}
where $x_{\mathrm{poly}}$ denotes the polymer or hybrid core class, $r_{\mathrm{core/drug}}$ the core-to-drug ratio, $\ell_{\mathrm{lig}}$ ligand density, $f_{\mathrm{PEG}}$ PEG fraction, $x_{\mathrm{solv}}$ the solvent system, $\omega_{\mathrm{mix}}$ the mixing rate, and $x_{\mathrm{pur}}$ the purification route. Categorical components are encoded by controlled vocabularies or one-hot/embedding representations, while continuous components are stored with units and uncertainty. Context variables $\bm{c}$ include release medium, temperature, serum content, cell line, passage number, dose, and exposure time. Candidate batches would be evaluated against a predefined target profile: hydrodynamic diameter between 80 and 150~nm, polydispersity index below 0.20, release half-life in a specified therapeutic window, cell viability above 80\% at the screening dose, sufficient uptake relative to a control, and an estimated synthesis-success probability above a pilot-defined threshold.

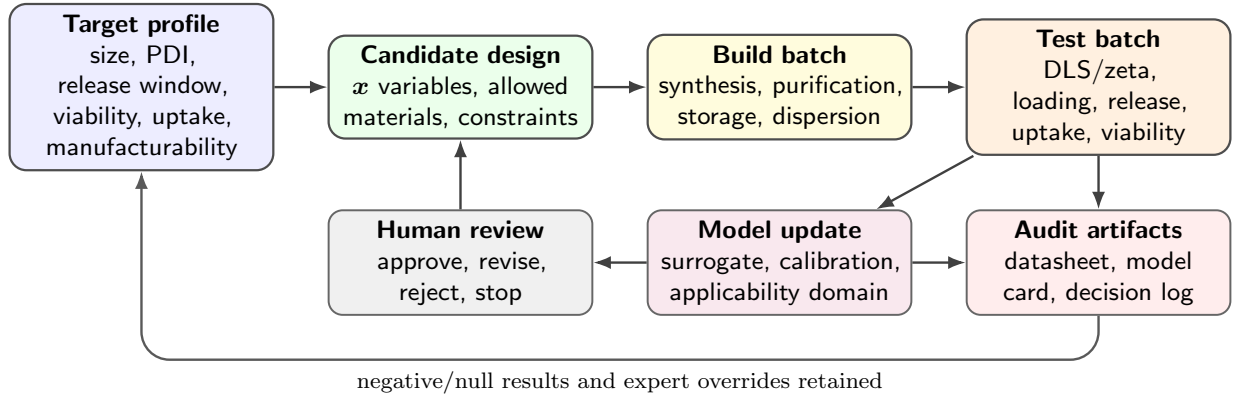
\begin{figure}[!htbp]
\centering
\resizebox{0.98\textwidth}{!}{%
\begin{tikzpicture}[
    font=\sffamily\footnotesize,
    stage/.style={rectangle, rounded corners=6pt, draw=black!70, line width=0.8pt, fill=#1, align=center, text width=3.05cm, minimum height=1.15cm},
    artifact/.style={rectangle, rounded corners=5pt, draw=black!60, line width=0.7pt, fill=#1, align=center, text width=3.05cm, minimum height=0.95cm},
    arrow/.style={-{Latex[length=2.2mm]}, line width=0.85pt, draw=black!75},
    feedback/.style={-{Latex[length=2.2mm]}, line width=0.85pt, draw=black!65, rounded corners=10pt}
]
\node[stage=blue!7]   (target)  at (0,0) {\textbf{Target profile}\\size, PDI, release window, viability, uptake, manufacturability};
\node[stage=green!8]  (design)  at (4.0,0) {\textbf{Candidate design}\\$\bm{x}$ variables, allowed materials, constraints};
\node[stage=yellow!16](build)   at (8.0,0) {\textbf{Build batch}\\synthesis, purification, storage, dispersion};
\node[stage=orange!12](test)    at (12.0,0) {\textbf{Test batch}\\DLS/zeta, loading, release, uptake, viability};

\node[artifact=purple!9] (model) at (8.0,-2.2) {\textbf{Model update}\\surrogate, calibration, applicability domain};
\node[artifact=gray!12] (review) at (4.0,-2.2) {\textbf{Human review}\\approve, revise, reject, stop};
\node[artifact=red!7]   (audit)  at (12.0,-2.2) {\textbf{Audit artifacts}\\datasheet, model card, decision log};

\draw[arrow] (target) -- (design);
\draw[arrow] (design) -- (build);
\draw[arrow] (build) -- (test);
\draw[arrow] (test) -- (model);
\draw[arrow] (model) -- (review);
\draw[arrow] (review.north) -- (design.south);
\draw[arrow] (test) -- (audit);
\draw[arrow] (model) -- (audit);
\draw[feedback] (audit.south) -- ++(0,-0.55) -| node[pos=0.25, below, font=\scriptsize] {negative/null results and expert overrides retained} (target.south);
\end{tikzpicture}}
\caption{Closed-loop drug-delivery nanoparticle pilot. The platform selects candidate batches under predefined constraints, the laboratory measures physicochemical and biological endpoints, the model is updated with complete metadata and negative results, and a human reviewer approves each next iteration.}
\label{fig:drugdeliverypilot}
\end{figure}

Figure~\ref{fig:drugdeliverypilot} summarizes the closed-loop pilot workflow. The figure separates the design variables, experimental measurements, model update, human review, and audit artifacts so that the pilot can be evaluated as a reproducible decision process rather than as an informal optimization exercise.

\begin{table}[htp!]
\centering
\caption{Example AI-ready record for the nanomaterial drug-delivery pilot. The values are illustrative; the purpose is to define the record structure required for a testable closed-loop study.}
\label{tab:workedpilot_record}
\footnotesize
\begin{tabularx}{\textwidth}{p{3.0cm}YY}
\toprule
\textbf{Record element} & \textbf{Example required fields} & \textbf{Use in model or governance layer} \\
\midrule
Material and batch identity & Polymer or inorganic core, drug identity, ligand, supplier, lot, batch ID, parent sample & Links all measurements and prevents mixing nominally similar but physically different samples \\
Synthesis and processing & Solvent, mixing rate, temperature, purification route, storage time, sterilization or filtration step & Defines process--structure relationships and enables failure-mode analysis \\
Physicochemical characterization & DLS size, polydispersity index, zeta potential, morphology, drug loading, uncertainty, replicate count, raw-file links & Provides model features and uncertainty-weighted labels for active learning \\
Assay context & Medium, serum fraction, temperature, cell line, passage number, dose, exposure time, controls & Makes uptake, release, and toxicity endpoints comparable across experiments \\
Outcomes and constraints & Release half-life, uptake signal, viability, colloidal stability, synthesis success/failure, excluded outliers with rationale & Defines objectives, feasibility constraints, and negative-result evidence \\
Governance and audit & Data-use status, responsible reviewer, model version, acquisition score, human decision, experiment date & Supports traceability and prevents uncontrolled use of restricted or incomplete records \\
\bottomrule
\end{tabularx}
\end{table}

The first closed-loop study could begin with 40--60 historical or newly curated batches and compare three strategies under the same experimental budget: expert-only selection, random or Latin-hypercube selection, and \platform{}-guided constrained acquisition using Eq.~\eqref{eq:acquisition}. In each iteration, the platform would recommend a diversified batch of, for example, four to six candidates; the domain expert would approve, modify, or reject the batch; the laboratory would measure size, zeta potential, release kinetics, uptake, and viability; and the resulting records would be appended to the dataset with negative and null outcomes preserved. The primary endpoint would be the number of experiments required to identify at least one candidate satisfying all predefined constraints. Secondary endpoints would include calibration error for toxicity or viability prediction, fraction of complete metadata fields, number of expert overrides, and whether the final recommendation lies inside the stated applicability domain.

A successful pilot would not need to prove clinical benefit. It would need to show that \platform{} improves preclinical materials decision-making relative to a fixed baseline while producing a complete datasheet, model card, and decision log. This benchmark directly addresses the concern that the manuscript should move beyond a general platform vision toward a concrete and falsifiable demonstration pathway.

\subsection{Materials Selection for Biomedical Implants}
Biomedical implants require long-term mechanical, chemical, and biological compatibility. A candidate implant material must be evaluated not only for stiffness, fatigue resistance, wear, and corrosion, but also for osseointegration, immune response, sterilization compatibility, and manufacturing reproducibility. Late-stage failure can be expensive and ethically problematic because the cost includes animal studies, clinical risk, and regulatory delay. Biocompatibility is therefore not a single material property; it is a context-dependent system response involving material, surface, host tissue, loading environment, and time \citep{Williams2008Biocompatibility}.

A platform workflow for implants would connect alloy or polymer composition, processing history, surface treatment, topography, coating chemistry, mechanical test data, corrosion data, in vitro biological response, and clinical outcome metadata. The knowledge graph could link surface chemistry to protein adsorption, protein adsorption to cell adhesion, and cell adhesion to osseointegration outcomes. Multi-objective optimization would then identify Pareto-optimal candidates rather than a single material ranked by mechanical strength alone. A strong pilot would compare platform-selected candidates against expert-only selections using fatigue-test failures, adverse biological responses, or number of screening cycles required to meet predefined mechanical and biological thresholds.

The pilot should distinguish between \emph{bulk material properties} and \emph{interface-controlled biological response}. For example, modulus and fatigue strength may be dominated by alloy composition, heat treatment, porosity, and manufacturing route, while protein adsorption, cell adhesion, bacterial colonization, and immune response may be dominated by surface chemistry, roughness, oxide composition, and sterilization history. This separation is essential because a model trained only on nominal composition cannot reliably predict implant performance.

\subsection{Photocatalytic Materials for Environmental and Biomedical Interfaces}
Photocatalytic materials are relevant to environmental remediation, antimicrobial surfaces, sterilization interfaces, and biomedical coatings. Their performance depends on optical absorption, charge separation, surface chemistry, defect states, stability, and interaction with surrounding media. AI-assisted screening can combine computed electronic descriptors with experimental activity measurements and environmental conditions to identify candidates with high efficiency and low toxicity.

In this use case, \platform{} would integrate descriptors such as band gap, band-edge positions, defect formation energies, adsorption energies, synthesis variables, illumination conditions, and medium chemistry with experimental photocatalytic activity, degradation pathways, microbial response, and cytotoxicity. The platform would not simply maximize activity, because a highly active photocatalyst may be unstable, toxic, or difficult to manufacture. Instead, the relevant objective is a constrained activity-to-risk profile: high degradation or antimicrobial rate, acceptable cytotoxicity, stable repeated cycling, and feasible processing. This use case is especially valuable for demonstrating multi-fidelity learning because inexpensive simulations can narrow candidate space before costly safety and stability assays.

A useful benchmark for this use case is not only the maximum degradation rate, but the stability of that rate over repeated cycles and the absence of harmful byproducts or leached species. The platform should therefore store illumination spectrum, photon flux, reactor geometry, catalyst loading, medium composition, target contaminant or microorganism, and analytical method. Without these fields, photocatalytic performance values are difficult to compare across laboratories.

\subsection{Personalized Biomaterials and Genomic Context}
This use case remains within the boundary of use stated in Table~\ref{tab:useboundary}: \platform{} may support exploratory and preclinical materials research, including cohort-aware stratification, but it is not proposed as a tool for patient-specific clinical prescription, autonomous intervention selection, or regulated clinical decision support. Genomic or phenotypic information should therefore be treated as governed research context rather than as a direct basis for patient-level recommendations.

Personalized or cohort-aware biomaterials create the most demanding integration problem. If an implant, scaffold, or drug-delivery system is to be studied for a patient subgroup, the platform must connect materials descriptors to governed biological information, such as genotype category, phenotype, immune status, anatomy, disease state, and treatment history. Genomic resources and protein-structure models can provide biological context, but their use in materials design requires careful governance and validation \citep{Sayers2025GenBank,Jumper2021AlphaFold,Abramson2024AlphaFold3}.

The platform should therefore support privacy-preserving analysis, controlled access, and federated or distributed learning when raw patient data cannot be centralized. Federated learning has been proposed as a way to train health AI models across institutions while reducing barriers to data sharing, and multicenter demonstrations show its potential for clinical prediction under distributed governance \citep{Rieke2020Federated,Dayan2021Federated}. In a personalized-biomaterials workflow, clinical collaborators could provide de-identified constraints, risk categories, or locally trained model outputs rather than raw identifiable records. The materials team would then optimize within those constraints while preserving an auditable separation between protected clinical data and materials-design decisions.

The appropriate near-term goal is cohort-aware materials research rather than individualized clinical prescription. The platform could test whether a material response differs across predefined, de-identified biological contexts while preventing direct re-identification and avoiding unsupported patient-level claims. Any transition from research stratification to clinical decision support would require separate validation, regulatory review, and monitoring.

\section{Platform Metadata and Minimum Information Requirements}
A common failure mode in cross-domain AI is that records are technically available but scientifically incomplete. Table~\ref{tab:metadata} proposes a minimum information structure for AI-ready biomedical materials data. The structure is intentionally broader than a conventional materials database because translational biomedical use requires assay context, safety information, and governance metadata.

\begin{table}[htp!]
\centering
\caption{Minimum information elements for AI-ready biomedical materials records.}
\label{tab:metadata}
\footnotesize
\begin{tabularx}{\textwidth}{p{2.55cm}YY}
\toprule
\textbf{Metadata class} & \textbf{Representative fields} & \textbf{Why it matters} \\
\midrule
Material identity and lineage & Composition, molecular structure, phase, supplier, lot, batch, parent sample, derived specimen, sample identifier & Enables reproducibility and links physical samples, process history, and digital records \\
Synthesis and processing & Precursors, temperatures, atmosphere, time, purification, surface treatment, sterilization, protocol version & Captures process--structure--property relationships and prevents protocol drift \\
Characterization and statistics & Instrument, calibration, method, raw files, uncertainty model, replicate count, statistical power, quality flags & Prevents mixing incompatible measurements and supports uncertainty-aware learning \\
Biological context & Cell line, organism, tissue, medium, dose, exposure time, endpoint, controls, human/animal-subject status & Makes biological response labels interpretable, ethical, and reusable \\
Performance endpoints & Mechanical, chemical, optical, release, toxicity, immune response, degradation, efficacy & Defines multi-objective optimization targets \\
Negative and null results & Failed synthesis, no activity, adverse response, low reproducibility, excluded outliers with rationale & Reduces publication and survivorship bias in model training \\
Governance metadata & Consent status, de-identification, access level, license, data-use limits, reuse restrictions, audit trail & Supports privacy, regulatory compliance, and responsible reuse \\
Model linkage & Feature pipeline, model version, prediction date, validation report, decision rationale, human override & Enables auditability and post hoc review of platform decisions \\
\bottomrule
\end{tabularx}
\end{table}

\section{Implementation Roadmap}
A realistic implementation should proceed in phases rather than attempting a fully autonomous platform at launch. Table~\ref{tab:roadmap} provides a staged roadmap.

\begin{table}[htp!]
\centering
\caption{Phased implementation roadmap for \platform{}.}
\label{tab:roadmap}
\small
\begin{tabular}{P{1.2cm}P{1.75cm}P{2.4cm}P{4.0cm}P{3.1cm}}
\toprule
\textbf{Phase} & \textbf{Indicative timing} & \textbf{Focus} & \textbf{Major activities} & \textbf{Exit criteria} \\
\midrule
I & 0--6 months & Foundation and governance & Define schemas, metadata, access controls, data-use agreements, and initial use cases & Versioned schema, governance plan, pilot datasets, initial dashboard \\
II & 6--12 months & Data integration and curation & Ingest public and local datasets; harmonize units; document provenance; build knowledge graph & Curated multimodal database with quality flags and queryable relationships \\
III & 12--24 months & Baseline models and validation & Train property, toxicity, and feasibility models; benchmark against held-out data; calibrate uncertainty & Model cards, validation reports, uncertainty calibration, baseline performance metrics \\
IV & 24--36 months & Closed-loop pilot & Deploy active learning for one or two use cases; run human-in-the-loop experiment selection & Demonstrated improvement over expert-only or random baseline \\
V & 36+ months & Translation and scaling & Add manufacturing, regulatory, and collaboration workflows; support controlled biomedical data & Auditable decision logs, user training, external partnerships, sustainability plan \\
\bottomrule
\end{tabular}
\end{table}

The first phase should not be dominated by model development. A platform with weak governance and poor metadata will produce fragile models regardless of algorithmic sophistication. Early effort should therefore prioritize data standards, provenance, stakeholder alignment, and pilot use cases. Once a reliable data foundation exists, baseline models can be trained and evaluated. Closed-loop experimentation should begin with constrained tasks where the cost of error is low and the value of rapid feedback is high.

Two workstreams should run in parallel. The \emph{technical workstream} builds schemas, adapters, models, dashboards, and workflow automation. The \emph{governance workstream} defines access tiers, data-use agreements, validation criteria, change-control procedures, and user training. A project should not advance from one phase to the next unless both workstreams meet their exit criteria. This prevents a common failure mode in AI platforms: rapid model development without sufficient data quality, user trust, or compliance readiness.

\section{Evaluation Metrics and Success Criteria}
The platform should be evaluated at three levels: data infrastructure, model performance, and translational workflow. Table~\ref{tab:metrics} lists representative metrics. These metrics are designed to avoid a common pitfall: judging an AI platform only by predictive accuracy. Accuracy matters, but a discovery platform must also improve data quality, reduce experimental burden, increase reproducibility, and support responsible decisions.

\begin{table}[htp!]
\centering
\caption{Representative metrics for evaluating platform performance. Quantitative targets should be customized for each pilot and compared against a fixed baseline.}
\label{tab:metrics}
\footnotesize
\begin{tabular}{P{2.4cm}P{3.35cm}P{2.75cm}P{4.0cm}}
\toprule
\textbf{Evaluation dimension} & \textbf{Representative metrics} & \textbf{Example pilot target} & \textbf{Interpretation} \\
\midrule
Data readiness & Fraction of records with complete required metadata; unit consistency; uncertainty reporting; negative-result capture & $\geq$90\% required fields complete for pilot records & Measures whether the platform is AI-ready rather than only data-rich \\
Interoperability & Number of mapped schemas; successful API calls; cross-domain query success rate; identifier resolution rate & Cross-domain query success for materials, assay, and governance fields & Measures whether data can move across materials and biomedical domains \\
Model performance & Prediction error, calibration error, confidence interval coverage, applicability-domain detection & Better calibrated uncertainty than an uncalibrated baseline model & Measures predictive reliability and uncertainty quality \\
Closed-loop efficiency & Improvement per experiment, experiments to target, reduction in failed trials, cost per validated candidate & 30--50\% fewer experiments to reach predefined pilot target & Measures whether the platform accelerates discovery \\
Scientific interpretability & Feature stability, mechanism consistency, expert review score, explanation reproducibility & Expert review confirms plausible drivers for top-ranked candidates & Measures whether model outputs are scientifically usable \\
Governance and compliance & Access violations, audit completeness, documentation completeness, model-change traceability & Complete audit trail for every model-generated recommendation & Measures whether platform operation is responsible and auditable \\
Adoption and sustainability & Active users, repeat workflows, training completion, partner contributions, maintenance cost & Repeat use by at least two independent pilot teams & Measures whether the platform can persist beyond pilot funding \\
\bottomrule
\end{tabular}
\end{table}

For the first nanomaterial pilot in Section~\ref{subsec:workedpilot}, the minimum success package should be defined before the first active-learning iteration. A practical package would include: (i) at least 90\% completeness for required metadata fields; (ii) a model card for every model used to rank candidates; (iii) an expected calibration error target for classification-style safety or viability predictions; (iv) a predefined baseline comparison using the same experimental budget; and (v) a complete audit trail for each recommendation, expert override, and model update. These mandatory metrics convert \platform{} from a broad architecture into a falsifiable pilot study.

For pilot studies, several metrics can be made explicit. Metadata completeness can be measured as
\begin{equation}
    C_{\mathrm{meta}}=
    \frac{1}{N}\sum_{i=1}^{N}
    \frac{\sum_{j=1}^{J} \mathbb{I}(m_{ij}\ \mathrm{present})}{J},
    \label{eq:metacomplete}
\end{equation}
where $m_{ij}$ denotes required metadata field $j$ for record $i$. Closed-loop acceleration can be measured relative to a baseline as
\begin{equation}
    A_{\mathrm{loop}}=
    \frac{E_{\mathrm{baseline}}-E_{\mathrm{platform}}}{E_{\mathrm{baseline}}},
    \label{eq:loopacceleration}
\end{equation}
where $E$ is the number of experiments required to reach a predefined target. For classification or risk models, uncertainty calibration should be evaluated explicitly, for example using expected calibration error,
\begin{equation}
    \mathrm{ECE}=\sum_{b=1}^{B}\frac{|B_b|}{N}
    \left|\mathrm{acc}(B_b)-\mathrm{conf}(B_b)\right|.
    \label{eq:ece}
\end{equation}
These metrics are simple, but they force the platform to demonstrate data readiness, decision efficiency, and calibrated confidence rather than only reporting a high cross-validation score.

A particularly important benchmark is comparison against a fixed expert-only baseline. For example, in a nanomaterial drug-delivery pilot, the platform could be evaluated by the number of experiments required to identify candidates meeting predefined release, toxicity, and manufacturability criteria. In an implant-materials pilot, it could be evaluated by the reduction in failed mechanical/biological screening cycles. Such benchmarks make the value of AI measurable and prevent vague claims of acceleration.

\section{Deployment Challenges and Mitigation Strategies}

Deployment risk is not a single technical problem. It arises from the interaction of incomplete data, model uncertainty, user behavior, privacy requirements, regulatory expectations, and operational cost. A useful operational risk score for a proposed workflow can be written as
\begin{equation}
    R_{\mathrm{workflow}} =
    r_{\mathrm{data}} + r_{\mathrm{model}} + r_{\mathrm{privacy}} +
    r_{\mathrm{reg}} + r_{\mathrm{impact}},
    \label{eq:risk}
\end{equation}
where the terms represent data-quality risk, model-generalization risk, privacy/security risk, regulatory risk, and consequence-of-error risk. Low-risk workflows may be appropriate for exploratory screening, whereas high-risk workflows require stronger validation, governance review, and human oversight.

\subsection{Interoperability with Existing Systems}
Many laboratories and companies already use legacy databases, spreadsheets, instrument-specific files, electronic laboratory notebooks (ELNs), laboratory information-management systems (LIMS), imaging repositories, and proprietary software. A successful platform cannot require all users to abandon existing workflows immediately. It should instead provide modular adapters, schema mapping, and staged migration. Standards such as OPTIMADE demonstrate the value of common APIs in materials data exchange \citep{Andersen2021OPTIMADE}. A similar adapter-based approach should be used for biomedical and laboratory data.

Interoperability should be non-destructive and bidirectional. The platform should ingest native files and local identifiers, map them into the minimum AI-ready schema, and preserve a link back to the original source record. Each adapter should record field-level provenance, unit conversions, missing-value flags, controlled-vocabulary mappings, and access restrictions. This approach allows laboratories to keep their operational systems while exposing the subset of information required for cross-domain learning, audit, and reproducible comparison. A practical deployment sequence is therefore to begin with read-only ingestion from spreadsheets and ELN/LIMS exports, then add validated write-back or dashboard functions only after schema mappings and governance controls have been tested.

\subsection{Scalability and Computational Cost}
The platform must support large datasets, high-dimensional imaging and spectroscopy, simulation workflows, and model training. Cloud infrastructure can provide elasticity, while high-performance computing may be needed for large simulations and training. However, unlimited scaling is not a strategy. Cost-aware acquisition functions, efficient surrogate models, data compression, and tiered storage should be integrated from the beginning. Equation~\eqref{eq:acquisition} explicitly includes experimental cost because scientific value must be balanced against time, money, and resource use.

\subsection{Model Reliability and Generalization}
Models trained on one laboratory, assay, material class, or patient population may not generalize to another. This is a major risk in cross-domain platforms. Mitigation strategies include external validation, domain adaptation, uncertainty calibration, out-of-distribution detection, and explicit applicability-domain reporting. Documentation practices such as model cards should be required before a model is used for decision support \citep{Mitchell2019ModelCards}. In high-stakes biomedical contexts, interpretable models and mechanistically constrained models should be prioritized when possible \citep{Rudin2019Interpretable}.

A key mitigation is an explicit applicability-domain report. For every recommended candidate, the platform should state whether the candidate lies inside the training distribution for composition, processing, assay context, and biological endpoint. Recommendations outside the training domain should be flagged as exploratory hypotheses rather than high-confidence predictions.

\subsection{Security, Privacy, and Regulatory Governance}
Biomedical integration creates privacy and compliance obligations that are stronger than those in many conventional materials databases. The platform should follow privacy-by-design principles, including least-privilege access, encryption, audit logs, de-identification, and formal data-use agreements. HIPAA and GDPR provide important regulatory reference points, while the NIST AI Risk Management Framework provides a broader structure for trustworthy AI risk management \citep{HHSPrivacyRule,EuropeanUnion2016GDPR,NIST2023AIRMF}. Federated or distributed learning should be considered when cross-institutional biomedical learning is valuable but raw data cannot be pooled \citep{Rieke2020Federated,Dayan2021Federated}.

The regulatory status of \platform{} would depend on intended use. A platform used for exploratory materials discovery is not automatically a regulated medical device. However, if platform outputs are used to support clinical decisions, generate evidence for a medical device, or guide adaptive software functions, FDA AI/ML guidance and predetermined change-control concepts should inform model validation, version locking, monitoring, and documentation \citep{FDA2021AIMLSaMD,FDA2025PCCP}. This distinction should be made explicit in deployment plans so that exploratory research, preclinical evidence generation, and clinical decision support are governed at appropriate risk levels.

\subsection{User Adoption and Workforce Development}
A technically sophisticated platform will fail if researchers cannot trust it or use it effectively. Adoption requires user-centered design, clear documentation, training modules, example workflows, helpdesk support, and transparent demonstration of scientific value. Interdisciplinary training is particularly important because materials scientists must understand biomedical data constraints, while biomedical researchers must understand materials descriptors, processing variables, and AI model limitations.

\begin{table}[htp!]
\centering
\caption{Major deployment risks and mitigation strategies.}
\label{tab:risks}
\small
\begin{tabularx}{\textwidth}{p{2.55cm}YY}
\toprule
\textbf{Risk} & \textbf{Consequence} & \textbf{Mitigation strategy} \\
\midrule
Incomplete metadata & Non-reproducible models and misleading correlations & Required metadata templates, automated quality checks, curator review \\
Dataset bias & Poor generalization and unsafe recommendations & Datasheets, subgroup analysis, external validation, active collection of missing regions \\
Overreliance on black-box models & Low trust and weak scientific interpretation & Interpretable models, mechanistic constraints, expert review, explanation audits \\
Privacy breach & Legal, ethical, and reputational harm & De-identification, access control, encryption, audit logs, federated analysis when needed \\
Regulatory mismatch & Delayed translation or unusable evidence & Early regulatory mapping, change-control documentation, validation plans \\
High operational cost & Unsustainable platform operation & Modular deployment, cost-aware workflows, tiered storage, shared infrastructure \\
Low user adoption & Platform underuse despite technical success & Training, user-centered design, pilot success stories, stakeholder governance \\
\bottomrule
\end{tabularx}
\end{table}

\section{Discussion}
The proposed platform is best understood as infrastructure for AI-enabled scientific reasoning rather than as a single database or software product. Its value emerges from connecting materials properties, biological response, mechanistic simulation, experimental feedback, and translational constraints. In this sense, \platform{} can serve as a bridge between materials informatics and biomedical translation.

In compact form, the framework differs from a conventional database in three ways: provenance, uncertainty, negative results, and governance are treated as first-class scientific objects; discovery is framed as constrained multi-objective decision support rather than single-property ranking; and every recommendation is tied to an auditable record of evidence, model version, applicability domain, and human review. These design choices are necessary because biomedical materials must satisfy performance, safety, manufacturing, and compliance constraints simultaneously.

The most practical near-term strategy is to begin with a limited number of high-value use cases. Nanomaterial drug delivery and implant-materials selection are strong candidates because they naturally require cross-domain integration and have measurable outcomes. As the platform matures, additional modules can incorporate more advanced digital twins, federated learning, autonomous laboratories, and regulatory evidence generation.

Sustainability should be treated as a scientific design requirement rather than as an administrative issue. A credible platform will need long-term schema stewardship, curator effort, user training, software maintenance, cost monitoring, and community governance. Without these functions, metadata quality will decay, model cards will become stale, and users will lose trust in the recommendations even if the initial algorithms perform well. For this reason, the roadmap should include a governance board, a curator role, versioned schema releases, documented deprecation policies, and periodic revalidation of active models.

The framework also creates a useful distinction between \emph{AI for prediction} and \emph{AI for scientific decision support}. Prediction asks what a model estimates for a candidate. Decision support asks whether the candidate should be synthesized, tested, rejected, or deferred given uncertainty, cost, safety, governance, and scientific value. This distinction is central to biomedical translation because a high predicted property value is rarely sufficient to justify advancement. The proposed architecture therefore treats models as components of an auditable decision system rather than as autonomous arbiters.

\section{Limitations and Future Work}
Consistent with its intended role as a framework and roadmap article, this manuscript defines the architecture, record model, governance posture, and pilot protocol for \platform{} rather than reporting a deployed system. The absence of a benchmark dataset or prototype implementation is therefore a boundary of the present work, not a claim of platform validation. The next stage should define at least one public or controlled-access pilot dataset with complete materials descriptors, assay context, uncertainty fields, negative results, and governance metadata. A second limitation is cross-domain ontology alignment: materials descriptors, nanomaterial safety terms, biomedical ontologies, clinical terminologies, and regulatory concepts do not map one-to-one, so curator review and iterative community governance will be necessary.

Third, biomedical labels are often noisy, context-dependent, and non-transferable. A toxicity label measured for one cell line, dose, medium, and exposure time may not generalize to another biological context. Fourth, privacy-preserving learning is necessary but not sufficient: federated learning can reduce barriers to data sharing, but it introduces additional challenges in model governance, site heterogeneity, fairness, auditability, and security. Fifth, AI-generated recommendations may be useful for prioritization but should not replace mechanistic understanding, expert judgment, or experimental validation. Finally, regulatory acceptance of AI-generated evidence remains an evolving area. The practical next step is therefore a limited pilot, such as nanomaterial drug delivery or implant-material selection, with predefined metadata requirements, baseline comparisons, validation assays, and decision logs. A successful pilot should publish or internally archive three artifacts: a curated dataset with a datasheet, a reproducible modeling workflow with a model card, and a decision log showing how model recommendations changed experimental choices. These artifacts would allow the community to evaluate whether the platform improved scientific decisions rather than merely adding another database.

\section{Conclusion}
This paper develops a scientific framework for \platform{}, an AI-native platform concept in which artificial intelligence is embedded from the outset as a core design, learning, and decision-making engine, rather than added as a post hoc analysis tool, while also incorporating FAIR data principles and governance-aware mechanisms to connect materials discovery with biomedical translation. The central conclusion is that the field does not simply need larger isolated databases; it needs interoperable, provenance-aware, uncertainty-aware infrastructure that links materials data, biomedical data, AI models, and experimental feedback. By representing discovery as a constrained multi-objective optimization problem and embedding active learning into a design--build--test--learn cycle, \platform{} is proposed as a framework for converting fragmented datasets into experimentally actionable recommendations, subject to validation in domain-specific pilot studies. The comparison with existing resources, minimum metadata model, use-case specifications, evaluation metrics, implementation roadmap, and risk controls provide a foundation for building and testing a platform that is scientifically credible, translationally useful, and responsible by design. The next milestone is not full-scale deployment, but a carefully bounded pilot that demonstrates interoperable data capture, calibrated prediction, human-in-the-loop experiment selection, and transparent governance in one high-value biomedical-materials workflow.

\section*{Author Contributions}
D.-M. Mei: conceptualization, methodology, project administration, supervision, funding acquisition, and writing--original draft. K. Acharya, C. M. Adhikari, M. Adhikari, S. Aryal, K. Bhatta, S. Bhattarai, N. Budhathoki, D. Chakraborty, S. Chhetri, S. Choudhury, T. A. Chowdhury, B. Cui, S. Dhital, K.-M. Dong, A. Ghasemi, B. D. S. Gurung, H. Oli, S. A. Panamaldeniya, L. Pandey, A. Prem, M. M. Rana, L.-W. Wang, Y. Yang, M. Zhou, and Q. Zhou: materials-science, chemistry, physics, device, and experimental-use-case input; investigation of platform requirements; and writing--review and editing. B. V. Benson, R. D. Cruz, A. M. Castillo, E. Z. Gnimpieba, R. Gapuz, H. A. Hashim, R. I. Harry, K.-E. Hasin, M. K. Hassanzadeh, M. K. Jha, D. Kim, K.-C. Kong, B. Lama, A. Mahat, N. Maharjan, A. Majeed, M. M. Masud, K. S. Moore, A. Nawaz, R. Pandey, Z. Peng, R. Rizk, C. S. Tadi, G.-L. Yin, C.-X. Yu, and D. Zeng: AI/ML, biomedical, data-infrastructure, education, governance, computational, and translational perspectives; validation of conceptual scope; and writing--review and editing. All authors reviewed the manuscript and approved the submitted version.

\section*{Data Availability}
No new experimental datasets were generated in this manuscript. The paper presents a conceptual platform architecture and cites public data resources and peer-reviewed literature.

\section*{Acknowledgments}
This work was supported in part by NSF OIA 2427805, NSF OISE 1743790, NSF PHYS 2117774, NSF PHYS 2310027, NSF OIA 2437416, DOE DE-SC0024519, DE-SC0004768, and a research center supported by the State of South Dakota. 

\section*{Conflict of Interest}
The authors declare no conflict of interest related to this conceptual manuscript.

\bibliographystyle{unsrtnat}
\bibliography{references_revised_major}

\end{document}